%% file: ms.tex
\documentclass[9pt,sigconf,letterpaper]{acmart}

\usepackage{tikz}
\usepackage{amsmath}
\usepackage{xspace}
\usepackage{multirow}
\usepackage{booktabs}

\usepackage{array}
\newcolumntype{L}[1]{
  >{\raggedright\let\newline\\\arraybackslash\hspace{0pt}}m{#1}}
\newcolumntype{C}[1]{
  >{\centering\let\newline\\\arraybackslash\hspace{0pt}}m{#1}}
\newcolumntype{R}[1]{
  >{\raggedleft\let\newline\\\arraybackslash\hspace{0pt}}m{#1}}

\usepackage{graphicx,subcaption}

\newcommand{\one}{({\em i})\xspace}
\newcommand{\two}{({\em ii})\xspace}
\newcommand{\three}{({\em iii})\xspace}

\newcommand{\eg}{{\it e.g.,}\xspace}
\newcommand{\ie}{{\it i.e.,}\xspace}
\newcommand{\etal}{{\it et al.}\xspace}

\newcommand{\iot}{IoT\xspace}

\newcommand\s[1]{(\S\ref{s:#1})\xspace}

\newcommand\fscore{$F_1$ score\xspace}
\newcommand\fig[1]{Figure~\ref{fig:#1}\xspace}

\newcommand\note[2]{\color{#1}\bf #2}

\newcommand\diana[1]{{\note{magenta}{diana: #1}}}

\newcommand\st[1]{{\color{blue} #1}}
\newcommand\takeaway[1]{\emph{#1}}

\AtBeginDocument{%
  \providecommand\BibTeX{{%
    \normalfont B\kern-0.5em{\scshape i\kern-0.25em b}\kern-0.8em\TeX}}}

\renewcommand\footnotetextcopyrightpermission[1]{} 
\setcopyright{none}
\acmDOI{}

\acmISBN{}

\acmConference[]{}{}{}
\acmBooktitle{}

\begin{document}

\date{}

\title{The Case for Retraining of ML Models for IoT Device Identification at the Edge}

\author{Roman Kolcun}
\affiliation{
  \institution{Imperial College London}
}
\author{Diana Andreea Popescu}
\affiliation{
  \institution{University of Cambridge}
}
\author{Vadim Safronov}
\affiliation{
  \institution{University of Cambridge}
}
\author{Poonam Yadav}
\affiliation{
  \institution{University of York}
}
\author{Anna Maria Mandalari}
\affiliation{
  \institution{Imperial College London}
}
\author{Yiming Xie}
\affiliation{
  \institution{Imperial College London}
}
\author{Richard Mortier}
\affiliation{
  \institution{University of Cambridge}
}
\author{Hamed Haddadi}
\affiliation{
  \institution{Imperial College London}
}

\renewcommand{\shortauthors}{Kolcun, R. et al.}

\begin{abstract}
Internet-of-Things (IoT) devices are known to be the source of many security
problems, and as such they would greatly benefit from automated management. This
requires robustly identifying devices so that appropriate network security
policies can be applied. We address this challenge by exploring how to
accurately identify IoT devices based on their network behavior, using
resources available at the edge of the network.

In this paper, we compare the accuracy of five different machine learning models (tree-based
and neural network-based) for identifying IoT devices by using packet trace data
from a large IoT test-bed, showing that all models need to be updated over time
to avoid significant degradation in accuracy. In order to effectively update
the models, we find that it is necessary to use data gathered from the
deployment environment, \eg the household. We therefore evaluate our approach
using hardware resources and data sources representative of those that would be
available at the edge of the network, such as in an IoT deployment. We show that
updating neural network-based models at the edge is feasible, as they require
low computational and memory resources and their structure is amenable to being
updated. Our results show that it is possible to achieve device identification and
categorization with over 80\% and 90\% accuracy respectively at the edge.


\end{abstract}

%
%

\maketitle

\input{intro}

\input{related}

\input{models}

\input{classification}

\input{edge}

\input{conclusion}




\bibliographystyle{ACM-Reference-Format}
\bibliography{ms}

\appendix


\end{document}

%% file: intro.tex
\section{Introduction}

Internet-of-Things (\iot) devices are the source of a large number of security
threats, particularly in domestic deployments~\cite{Alrawi2019}. These devices
and their platforms would benefit from active and, in particular, automated
management -- but automating such tasks requires robustly identifying devices to
be able to apply appropriate policies, actions, and updates. In this
environment, the natural way of identifying \iot devices is to analyze their
network behavior at the home router: devices cannot hide behavior as, by
definition, they must interact over the network in order to provide
functionality. Performing analyses of network behavior at the home router is
robust in terms of privacy, scalability and not relying on dependencies from
manufacturer-provided cloud-services.  Furthermore, there is already nascent
support for summarizing such analyses' results via the MUD
standard~\cite{rfc:8520}. As such, several use cases for IoT device
identification arise, summarized in Table~\ref{table:use-cases}.

Previous work has resorted to machine learning to carry out \iot device
identification~\s{related}.  The usual approach entails training machine
learning models offline or in a cloud environment~\cite{Miettinen2017,
Hafeez2020, Sivanathan2018, Pashamokhtari2020}, and run inference to identify
the devices at the home routers. However, the training and validation of these
models is done on a particular set of devices, and for a limited time period. As
a consequence, these models have good accuracy only when the training and
inference is run on the same dataset. This means that static, pre-trained models
cannot be used for identification across different home \iot networks while
ensuring good accuracy. Moreover, different users have different usage patterns,
and their social behavior might evolve over time. For example, a user might
have a certain routine while working and consequently interaction with their IoT devices, 
but interact differently with the devices while he is on holiday at home. 
Another example might be watching more TV when they stay indoors due to different factors
(such as weather), while interacting less when the user is away from home for different reasons.
Thus, it is paramount to update the models to take into account the user behaviour.

In this paper, we first evaluate five machine learning models (Random Forest
Classifier, Decision Tree Classifier and three neural network-based models)  for
\iot device identification using data from a large scale test-bed with $43$
devices, and show that all models exhibit a degradation in accuracy over time.
While this degradation diminishes if the models are trained over a longer period
of time and if they are not used for prediction for an extensive period of time,
properly counteracting it requires regular retraining of the model. We achieve
this by updating the models with local test-bed data at the network gateway to
maintain their accuracy over time.  Model training can be centralized or
decentralized, \ie carried out in the cloud or at the edge. The former requires
that details of traffic dynamics from deployed devices are reported to some
central location. This is untenable due to both the scale of deployment
anticipated for \iot and the privacy concerns inherent in reporting such
traffic. As a result, we focus on training at the edge, where compute capability
is limited.  Based on our comparison of different models, we choose neural
network-based models for edge deployment, since these models can be updated.
Given the limited computational capabilities of available edge nodes, we show
how to further reduce the computational cost of model update by freezing layers
in the neural networks while quantifying the impact this step has on accuracy.
Our methodology for updating models in the gateway enables identification of
devices with over 80\% accuracy and device categories with over 90\% accuracy
(\fscore).

The main contributions of the paper are as follows:
\begin{itemize}
    \item We compare five different types of machine learning models for \iot
    device classification in terms of their accuracy and show that in all five
    cases the accuracy decays over time. We analyze the possibility of keeping
    the models updated using data specific to a household, and the resources required to update
    the models at the edge.
    \item We show that models need to be updated using data specific to the
    household. A model updated using data from one household does not perform
    well on another household and vice versa. 
    \item We demonstrate that updating the models through retraining or partial
    retraining at the edge of the network on a representative edge device
    (Raspberry Pi) is feasible.  
\end{itemize}

The remainder of the paper is organized as follows. We first provide an overview
of related work~\s{related}. We then present five different machine learning
models used in our work, our test-bed, and the set of experiments
conducted~\s{models}.  We show that the accuracy of the models decays over time,
but updating them improves device identification~\s{classification}. We
demonstrate how models can be updated at the edge, and how they perform for
device identification~\s{edge}. Finally, we discuss future
challenges~\s{discussion} and conclude our work~\s{conclusions}.

\begin{table}[]
\footnotesize
  \captionsetup{skip=0.2em, font=small}
  \caption{\label{table:use-cases} IoT device identification use cases}
    
    \begin{tabular}{L{0.25\linewidth} L{0.65\linewidth}}
    \toprule
    \bf Use case  & \bf Example \\ 
    \midrule
    Network management  & Prioritisation depending on device purpose (medical devices data take precedence over weather app data) \\
    \midrule
    Network security    & Apply device policy based on MUD profiles \\ 
    \midrule
    Behavioral analysis & Understand users behavior\\ 
    \bottomrule
    \end{tabular}
\end{table}

%% file: related.tex
\section{Related Work}
\label{s:related}

In the last decades, a vast number of machine learning-based network monitoring
and  Internet traffic classification techniques, both in a distributed and
centralized manner, have been explored~\cite{Moore2005, Pacheco2019,
Nguyen2008}. However, not all methods are suitable for \iot, and some of these
techniques are adopted and customized for \iot; therefore, in this section, we
focus only on techniques used for analyzing \iot traffic. 

\textbf{Traffic Classification for \iot.} Offline \iot network traffic analysis
is used for understanding various \iot device or user
behaviors~\cite{ren-imc19, Apthorpe2016, Tahaei2020}. For example,
Yadav~\etal~\cite{Yadav2019} studied traffic from a dozen \iot devices in a lab
environment to understand network service (\eg DNS, NTP) dependencies and
robustness of device function when connectivity is disrupted.
Apthorpe~\etal~\cite{Apthorpe2016} analyzed the traffic rates of four \iot
devices, showing that observations about user behavior can be inferred even
from encrypted traffic. Similarly, traffic categorization using both statistical
and machine learning techniques has been performed by
Amar~\etal~\cite{Yousef2018}.  

\textbf{Device Identification and Anomaly detection in \iot.} The \iot device
identification is the first step towards finding any malicious or unknown \iot
device in the network. Generally, many \iot
devices have a unique identifier assigned during manufacturing such as MAC
address or hardware serial numbers. Even though these unique addresses could
reveal some information about the device manufacturer, still the full
identification of malicious/abnormal devices in the network using only these
unique addresses is not possible. Thus, behavior-based \iot device
identification methods, which use traffic classification mechanisms have gained
attention recently~\cite{Meidan2017, Miettinen2017, Hafeez2020, Trimananda2020}.
The \iot applications, \eg anomaly detection and prediction, require low
latency and privacy at the edge, and traffic based behavior identification is
needed to be done in the real-time at the gateway level for security and data
privacy purpose~\cite{Yang2019, Magid2019,Kusupati2018}.  

\textbf{Machine Learning for Device Identification.}  Machine Learning in \iot
at the edge is still in its infancy, due to partly lack of available network
data in the wild and lack of compact machine learning models. The recent uptake
in resource-constrained machine learning~\cite{tflite,banbury2020, Painsky2019,
Feraudo2020} has led to a renewed interest in applying machine learning to \iot
network-related problems, specifically network traffic
classification~\cite{Ortiz2019, chimera2014, Feng2018}, anomaly
detection~\cite{Nguyen2019,Feraudo2020} and device
identification~\cite{Meidan2017, Miettinen2017, Hafeez2020, Sivanathan2018}.
Sivanathan \etal~\cite{Sivanathan2018} used multi-stage classifiers~(Naive Bayes
Multinomial and Random Forest classifier) for \iot device classification and
achieved accuracy from 99.28\% to 99.76\% with classifiers trained on 1 to 16 days
data from 28 unique \iot devices. Nguyen \etal\cite{Nguyen2019} trained Gated
Recurrent Network (GRU) for federated learning  for anomaly detection using 33
devices categorized in 27 categories and for evaluation, deployed 13 devices and
found only 5 are vulnerable to the Mirai attack when the attack is injected in the
local network. The attack was detected within 30 minutes. Many identification
works train machine learning models offline or in a cloud
environment~\cite{Miettinen2017, Hafeez2020, Sivanathan2018, Pashamokhtari2020}
and run inference to identify \iot devices on local gateways. The training and
evaluation is done only for a set of devices for a limited time period, thus
inference achieves a good accuracy when testing data is similar to the training
data.  However, for real world scenarios, a pre-trained model on a small set of
devices would not work on a large set of unknown \iot devices. The
identification accuracy may drop when the inference data is different from the
training dataset, therefore, requiring retraining of the model for the local
setup. None of the works above have looked at or addressed this problem,
therefore, in our work we investigate retraining requirements for maintaining
high identification accuracy over an extended period of time.

%% file: models.tex
\section{Dataset and Models}
\label{s:models}
In this section, we discuss the device types and categories in our test-bed, and the models used to classify them. Our evaluation focuses on two classification problems:

\begin{itemize}
\item device classification, \ie assigning the network flow to a particular
device that generated the flow;
\item category classification, \ie assigning the network flow to a category of
devices (\eg media, surveillance, or appliances).
\end{itemize}

For the purpose of evaluations we tagged all the network flows with a device ID
and a category ID and used these tags to train each model.  Our dataset consists
of data collected during $d_{max} = 21$~days. We denote the dataset as $D =
{d_i}, 1 \leq i \leq d_{max}$. Each model is trained on data collected for a
different number of consecutive days. We use sliding windows with different
window lengths $w \in \langle 1 \ldots 7 \rangle$ days.  Trained models are
evaluated on all days ahead of the training days. If a model is trained on a
dataset with a window length $w$ using data from days $\langle d_x \ldots
d_{x+w} \rangle, x+w < d_{max}$, then the model is evaluated on \emph{prediction
days} $p$, using data from days $d_{x+w+p}, 1 \leq p \leq p_{max}, x+w+p_{max}
\leq d_{max}$. For the evaluations, we use the upper limit of $p_{max} \leq 14$,
\ie we predict for the maximum length of two weeks. 

We create and evaluate two \emph{groups} of settings: \one one model for
\emph{all} devices/categories;\two one model \emph{per} device/category.  In
the first case, the model predicts to which device/category the given network
flow belongs, \ie the model performs multiclass classification. In the second
case, the model predicts the probability with which the network flow was
generated by the given device/category, \ie the model performs binary
classification. 

In the rest of this paper we adopt the following taxonomy: the model
\emph{group} refers to whether we use one model for all devices/categories or
one model per each device/category. The model \emph{type} refers to which
classifier was used, \eg Random Forest Classifier or Convolutional Neural
Network. 

In the case of a single model for all devices/categories (referred to as
\emph{all-device} or \emph{all-category} model) the inference part is rather
simple as the model performs multiclass classification directly to given
device/category. In the second case, a single model is created for each
device/category (referred to as \emph{per-device} or \emph{per-category}) and
performs binary classification. In order to find out which device/category
generated the network flow, inference on all models needs to be executed. The
model which predicts the device/category with the highest probability is chosen
as the output. Therefore, for a single classification we need to run as
many model inferences as the number of devices/categories. This could be further
optimized using hierarchical model evaluation~\cite{Pashamokhtari2020}.

\begin{table}[]
\footnotesize
  \captionsetup{skip=0.2em, font=small}
  \caption{\label{table:devices}
    Categorized \iot devices in our test-bed.
    }

    \begin{tabular}{L{0.25\linewidth} L{0.65\linewidth}}
      \toprule
      \bf Category & \bf Device Name\\
      \midrule
      Surveillance & Blink camera, Bosiwo camera, Spy camera, D-link camera,
      \st{Ring doorbell}, \st{Wansview camera}, Xiaomi camera, Yi camera\\
      \midrule
      Media           & Apple TV, Fire TV, \st{Roku TV}, Samsung TV\\
      \midrule
      Audio           & Allure speaker, Echodot, Echospot, Echoplus, \st{Google
      mini}, Google home\\
      \midrule
      Hub             & Insteon hub, Lightify hub, Philips hub, Sengled hub,
      Smartthings hub, \st{Xiaomi hub}, \st{Switchbot hub}\\
      \midrule
      Appliance       & Smart Kettle, Smarter coffee machine, Sousvide cooker, Xiaomi cleaner, Xiaomi rice cooker\\
      \midrule
      Home automation & Honeywell thermostat, Nest thermostat, Netatmo weather
      station, TP-link bulb, TP-link plug, Wemo plug, Xiaomi plug, Yale door
      lock, \st{LED strip}, \st{Smartlife remote}\\
      \bottomrule
    \end{tabular}
\end{table}

\subsection{Dataset and Experiments}
\label{s:dataset}

To capture data, we built two test-beds: \one the \emph{large test-bed}
that currently comprises 43 different \iot devices and \two the \emph{small
test-bed} that contains a subset of 9 devices from the large one. We selected
these devices to provide diversity within and between each category:
surveillance, media, audio, hub, appliance, and home automation devices.
Table~\ref{table:devices} describes the devices in our test-beds, by category.
Devices in blue are common for both test-beds. 

In the large test-bed, in addition to the devices, a Linux server running Ubuntu
18.04 with two Wi-Fi cards for 2.4~GHz and 5~GHz connections, plus two 1~Gbps
Ethernet connections for LAN and Internet connectivity are part of the setup.
The server sits outside of any firewall and has a public IPv4 address. However,
to match a regular home network environment, all \iot devices are behind a NAT
setup and cannot be accessed directly from the Internet. A similar setup but
with a Raspberry Pi Model 4 (4~GB RAM), with only one Wi-Fi card operating at
2.4~GHz and only one Ethernet connection is setup for the small test-bed. The
monitoring software automatically detects the connection of a new device to the
network, assigns it a local IP address, and starts capturing packets using
\emph{tcpdump}. Each device's traffic is filtered by MAC address into separate
files.

The \iot devices can usually be controlled via a \emph{companion device} such as
a smartphone application, an Alexa voice assistant, or a Google Home. Our test-bed
allows us to perform manual and automated experiments on the IoT devices using
these companion devices. In this case, the monitoring software captures the
network traffic of both \iot and companion device into separate PCAP files.
The test-bed allows us to capture several network traces for each device under
different conditions:

\textbf{Idle periods}. The devices are not actively used by automated
experiments, but they might be unknowingly activated by people present in the
lab (\eg motion or noise detected by a camera if a person is passing by) or
rarely checked by a researcher to see whether the device is still connected to the
Internet. 

\textbf{Automated experiments}. Automated experiments can be carried out using a
selected companion device from two Nexus 5X smartphones with Android 6.0.1, an
iPhone 5S, an Amazon Echo Spot, and a Google Home. Using the companion device,
our software automatically performs an action on the device such as switching
light bulbs on and off either via scripted Android Debug Bridge (ADB) interactions
with their control app (if the companion device is a smartphone) or synthesized
voice commands (if the companion device is an Amazon Echo Spot or Google Home).

We collected data from the \emph{idle period} of the large test-bed for the
duration of 21 days. Additionally, we collected data form the \emph{automated
experiments} from both test-beds for 7 days. A list of all possible interactions
with the test-beds was compiled and actions from this list were executed. The
experiments followed three patterns:

\begin{description}
\item[light] - simulating a single professional living alone. Several actions
were executed in the morning (\eg turning the lights on/off) and/or listening
to the news, then an action during the day (\eg checking the doorbell or a
home camera), and several actions in the evening (turning the lights on/off or
streaming TV from the Internet). 

\item[medium] - simulating a single professional working from home where in
addition to the \emph{light} pattern, the number of actions was higher and
several actions were added (\eg streaming a music using a smart speaker). 

\item[heavy] - simulating a family life over a weekend, where in addition to the
\emph{medium} the number of actions increases and smart speakers and smart TVs
were used more often to stream content from the Internet. 

\end{description}

All the actions performed on each test-bed were independent and the test-beds do
not communicate with each other.

\begin{center}
\begin{table}
\footnotesize
 \captionsetup{skip=0.2em, font=small}
\caption{\label{tab:features}
List of features used for training of ML models.}
\begin{tabular}{l l}
\toprule
\bf Feature Name         & \bf Feature Description  \\
\midrule
{\it src\_port}     & source port                                 \\   
{\it dest\_port}    & destination port                            \\
{\it bytes\_out}    & number of bytes sent                        \\
{\it bytes\_in}     & number of bytes received                    \\
{\it pkts\_out}     & number of packets sent                      \\
{\it pkts\_in}      & number of packets received                  \\
{\it ipt\_mean}     & mean of inter-packet interval               \\
{\it ipt\_std}      & standart deviation of inter-packet interval \\  
{\it ipt\_var}      & variance of inter-packet interval           \\
{\it ipt\_skew}     & skewness of inter-packet interval           \\
{\it ipt\_kurtosis} & kurtosis of inter-packet interval           \\
{\it b\_mean}       & mean of packet sizes                        \\
{\it b\_std}        & standard deviation of packet sizes          \\
{\it b\_var}        & variance of packet sizes                    \\
{\it b\_skew}       & skewness of packet sizes                    \\
{\it b\_kurtosis}   & kurtosis of packet sizes                    \\
{\it duration}      & duration of the stream                      \\
{\it protocol}      & protocol ID                                 \\
{\it domain}        & second and top level domain                 \\ 
\bottomrule

\end{tabular}
\end{table}
\end{center}
 
\subsection{Processing Traces}\label{s:traces}

All the network traffic from both test-beds is stored locally on the computer
acting as a router in a \texttt{pcap} format. These files are then processed by
\emph{joy} \cite{joy} utility which extracts the following features from each
TCP/UDP network flow (summarized in Table~\ref{tab:features}): source and
destination IP address, source and destination port number, number of packets
sent and received, bytes of packets sent and received, starting and ending time
of the flow. Additionally, \emph{joy} extracts DNS request and replies which can
be later analyzed. Flow features are extracted if the network flow is inactive
for more than ten seconds, or if the network flow is active for more than 30
seconds. If the network flows continues, a new record is created. It means, that
a set of features is extracted at latest after 30 seconds which allows us to
perform near on-line device classification. 

The extracted features contain also information about the first up to \emph{N}
packets. We used the default value of $N = 50$. This information includes data
about packet sizes and inter-packet intervals. Using information about packets,
additional features are computed, \ie duration of the flow, and for both, packet
sizes and inter-packet intervals, mean, standard deviation, variance, skew, and
kurtosis is computed. Each flow is assigned the device ID and the category ID. 

The list of DNS responses is used to map IP addresses to domain names. We chose
not to use IP addresses as a feature because they may not be consistent due to
the nature of the services running in cloud. A virtual server may migrate to
another physical server and its IP may change. Or a new server might be
temporarily started to balance the load. Additionally, many large manufactures
are using DNS load balancing where the same domain is translated to different IP
addresses. Therefore, we decided to use the domain name as a feature. However,
we noticed that many times the domain name differs on the third or further
level. This is especially common when a content delivery network is contacted.
Therefore, we decided to use only the second and top level domain name as a
feature.

Dataset from the \emph{idle} part of the large test-bed consisted of 6,452,100
network flows. Data collected during the \emph{active} experiments on the large
test-bed contained 4,691,596 flows (1,371,516 were from the devices common for
both test-beds) and 1,177,765 flows from the small test-bed.

\subsection{Model Types}

We have selected five different learning algorithms from classic (supervised) Machine
Learning (ML) and Neural Networks (NN) models to identify \iot \emph{devices}
and their \emph{categories}. The decision of which machine learning algorithms
to use is driven by memory footprint and inference time requirements.

\textbf{Classical ML algorithms.} We initially chose Decision Tree Classifier (DTC)
\cite{Safavian1991} and Random Forest Classifier (RFC) \cite{Breiman2001}, as,
according to previous work on network traffic analysis and classification
\cite{Pinheiro2019, Sivanathan2018}, these models showed the highest accuracy
among other widely-used classical ML methods (\eg Gaussian Naive Bayes,
K-Nearest Neighbors, or Support Vector Machines). 

DTC makes a prediction by moving each data point through a tree-like structure of
nodes and leaves. Each node in the decision tree contains conditions that will
be checked in order to classify a data point. Each data point is passed through
several decision nodes until it reaches the leaf which provides the final
classification. The accuracy of DTC can be improved by increasing the number of
decision nodes and adding more complex validation conditions. 

RFC consists of multiple DTCs where each decision tree makes data point
classification independently of others based on a randomly selected subset of
features. The class with the most votes becomes the RFC model prediction.

\textbf{Neural Networks (NN)}. We evaluated models based on three types of
neural networks: \one Fully-Connected (FC) NN \cite{Goodfellow-et-al-2016}, \two
Long Short-Term Memory (LSTM) networks \cite{Hochreiter1997}, and \three
Convolutional Neural Networks (CNN) \cite{Lecun1995}.

Fully-Connected (FC) NN are a type of feedforward networks that consist of
series of fully-interconnected layers. Fully-connected NN are
structure-agnostic, and thus applicable for analysis of any type of input data,
including network traffic. 

Long Short-Term Memory Networks (LSTM) are a type of Recurrent Neural Networks
(RNN). Similar to RNN, LSTM consist of a chain of repeating learning modules,
but the increased number of interacting layers (four instead of one) allow LSTM to
learn long-term dependencies more effectively. This feature makes LSTM useful
for traffic classification tasks where analysis of large long-term
data is required. For example, previous works \cite{Aceto2019, Lopez-Martin2017}
applied LSTM combined with CNN for mobile and IoT traffic classification.

In CNN, unlike FC NN, not all neurons from two adjacent layers are
interconnected. Having convolution layers allows CNN to activate specific
filters that are the most important for a given learning task on a given
intermediate layer. CNN are mostly used for classification of signals, images
and videos, but recently CNN have also been used for network and mobile traffic
classification tasks \cite{Wang2017, Wang2020}.

\textbf{Comparing Classical ML against NN models}. We found that the accuracy of
DTC and RFC models is slightly higher when compared to neural networks used in
our evaluation for IoT device classification. However, both classical ML and NN
models lose accuracy over time and thus require frequent updates with new
training data \s{classification}. This is particularly problematic with RFC and
DTC because their model sizes scale linearly with the number of the training
set and can easily reach hundreds of MB~\s{model-sizes}. Another disadvantage of
decision-tree-based algorithms (in their original form) is the inability to
update them with new data.  Therefore, as new training data are received, the
only viable solution is to merge them with the previously collected training
data and retrain the models on the whole dataset.  Although some modifications
of decision-tree-based algorithms allow incremental training
\cite{Lakshminarayanan2014, Saffari2009}, the observed model size in one of such implementations
\cite{Lakshminarayanan2014} tested by us was similarly large as the original RFC
models.

On the other hand, the advantage of NN is two-fold: \one the
size of the model is rather small (varies from hundreds of kilobytes to very few
megabytes) and \two it is possible to update them with new data, \ie there
is no need to keep historical data and new training data can be used to update
the current models. 

Detailed description of the models used throughout our evaluation is as follows:

\begin{description}

\item[RFC] - maximum depth of the tree was set to 100, minimum number of samples
per leaf was 1, minimum number of samples before a node can be split was set to 10, and the number of estimators was 3. 

\item[DTC] - maximum depth of the tree was set to 100, minimum number of samples
per leaf was 1, and the minimum number of samples before a node can be split was
set to 10. 

\item[Fully Connected (FC) NN] - consists of following layers (integer represents size of each layer): FC(32), FC(64),
FC(128), FC(256), Output Layer.

\item[LSTM] - consists of following layers: LSTM(200), LSTM(100), LSTM(50),
LSTM(25), Dropout(0.2), Output Layer.

\item[Conv1D] - is 1D CNN and consists of following layers: Conv1D(64, 3), Conv1D(64, 3),
Dropout(0.2), MaxPooling1D(), Flatten(), FC(100), Output Layer. 

\end{description}

\begin{figure*}[!bpt]
    \centering
    \begin{subfigure}[t]{.45\linewidth}
\includegraphics[width=\linewidth]{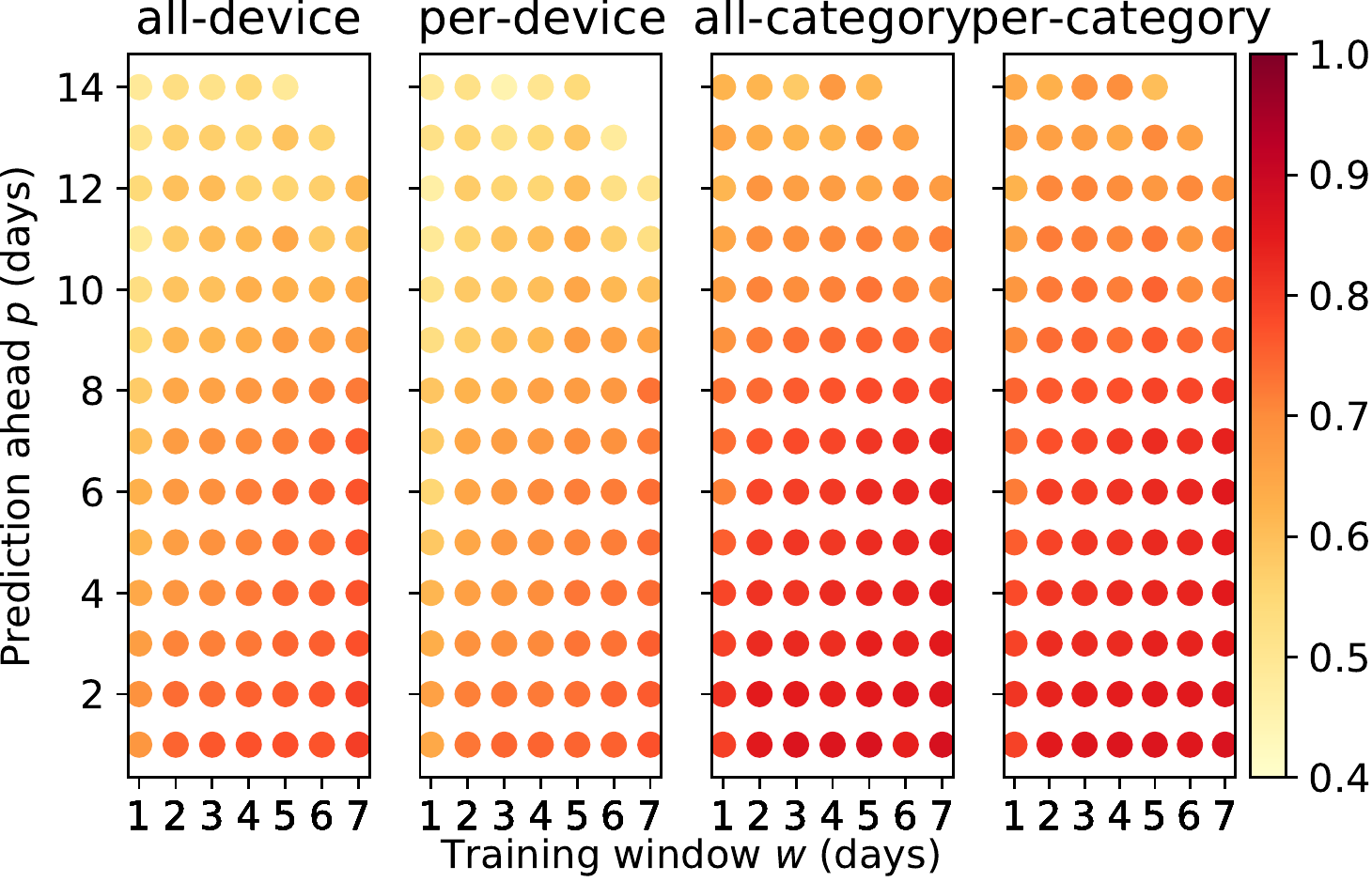}
    \caption{Fully connected model}
    \label{fig:testbed_pred_nn}
    \end{subfigure}
    \begin{subfigure}[t]{.45\linewidth}
\includegraphics[width=\linewidth]{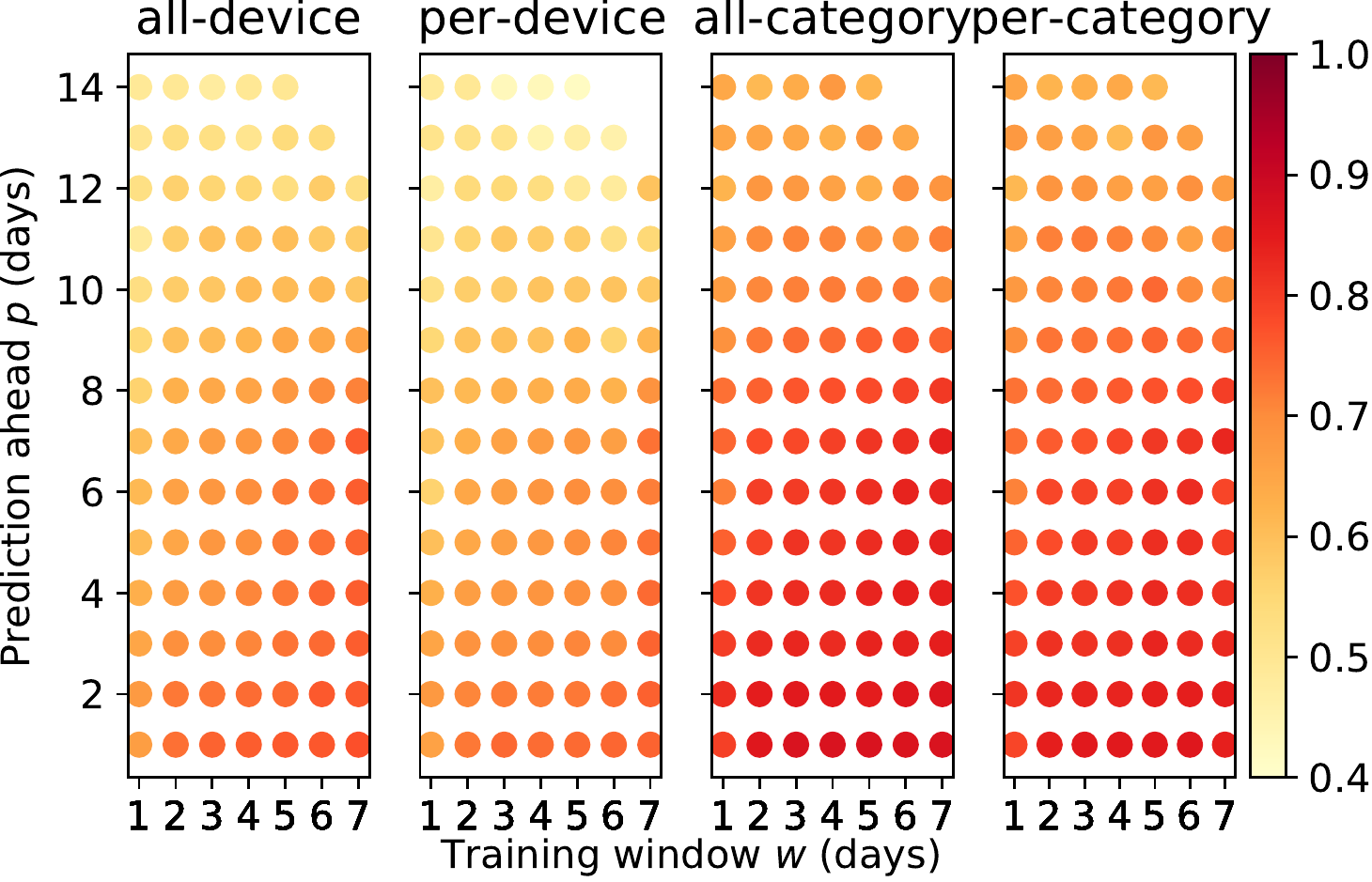}
    \caption{Long short-term memory model}
    \label{fig:testbed_pred_lstm}
    \end{subfigure}
    \begin{subfigure}[t]{.45\linewidth}
\includegraphics[width=\linewidth]{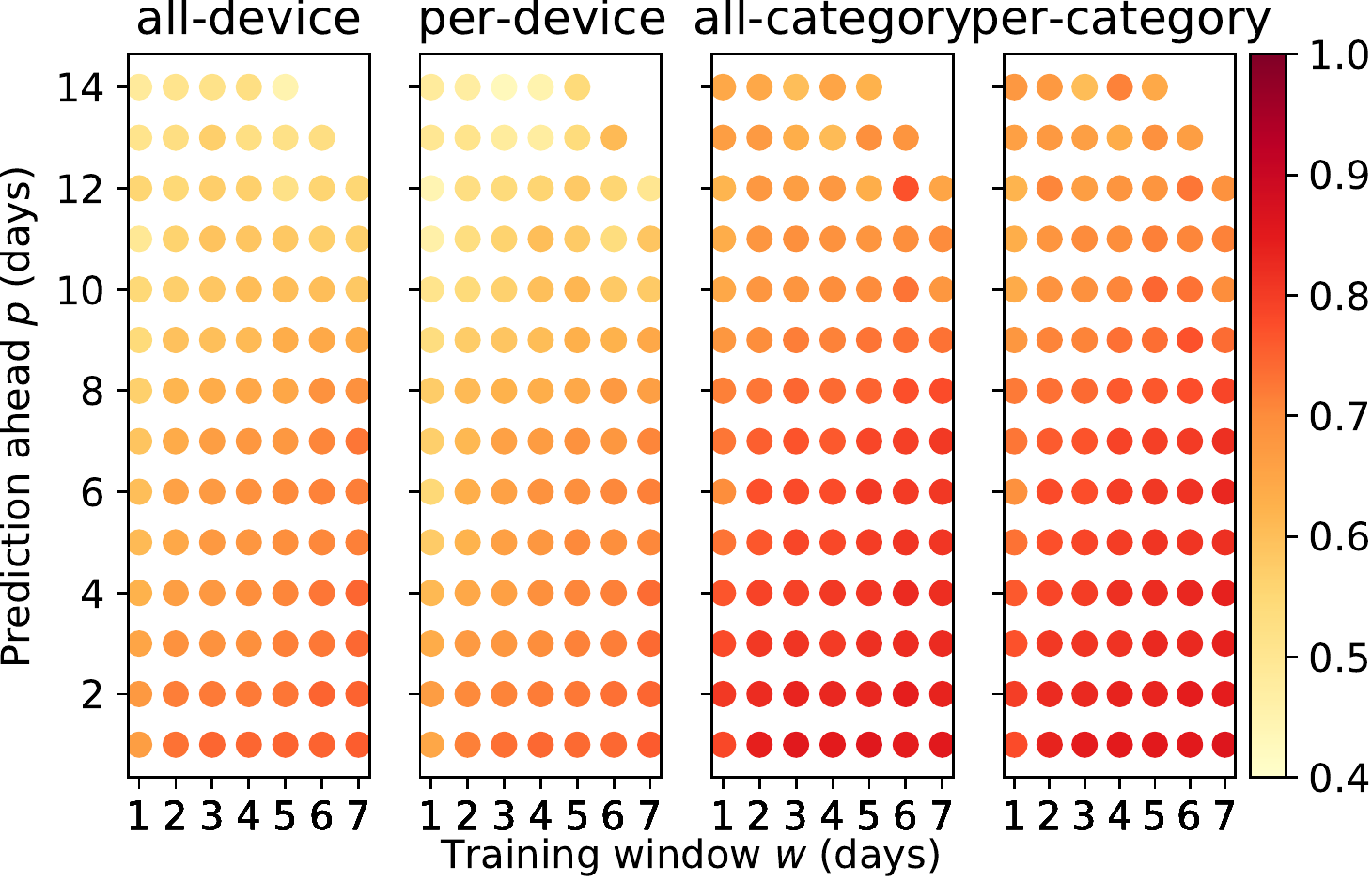}
    \caption{1D convolutional model}
    \label{fig:testbed_pred_conv1d}
    \end{subfigure}
    \begin{subfigure}[t]{.45\linewidth}
\includegraphics[width=\linewidth]{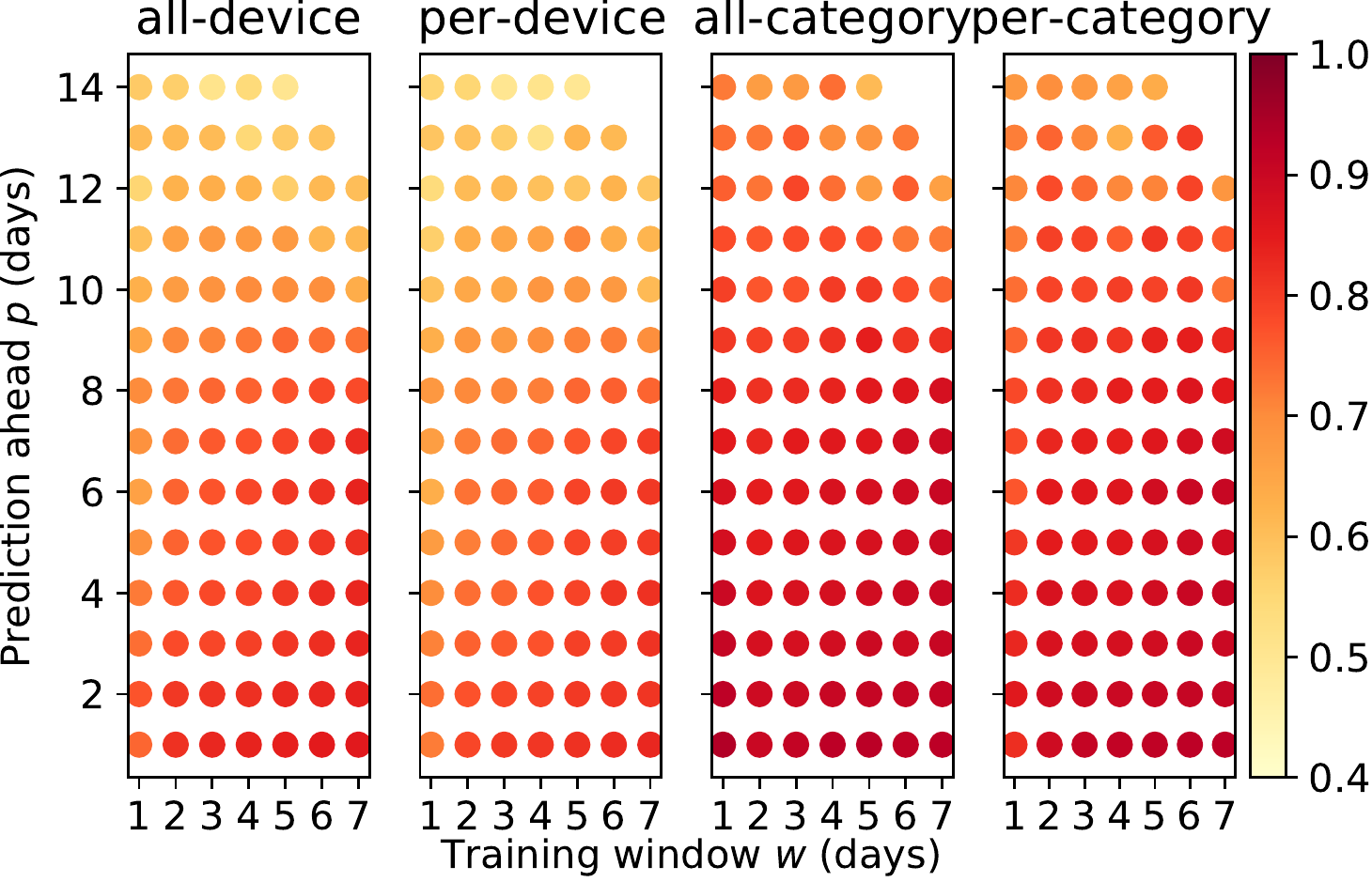}
    \caption{Random forest classifier}
    \label{fig:testbed_pred_rfc}
    \end{subfigure}
    \begin{subfigure}[t]{.45\linewidth}
\includegraphics[width=\linewidth]{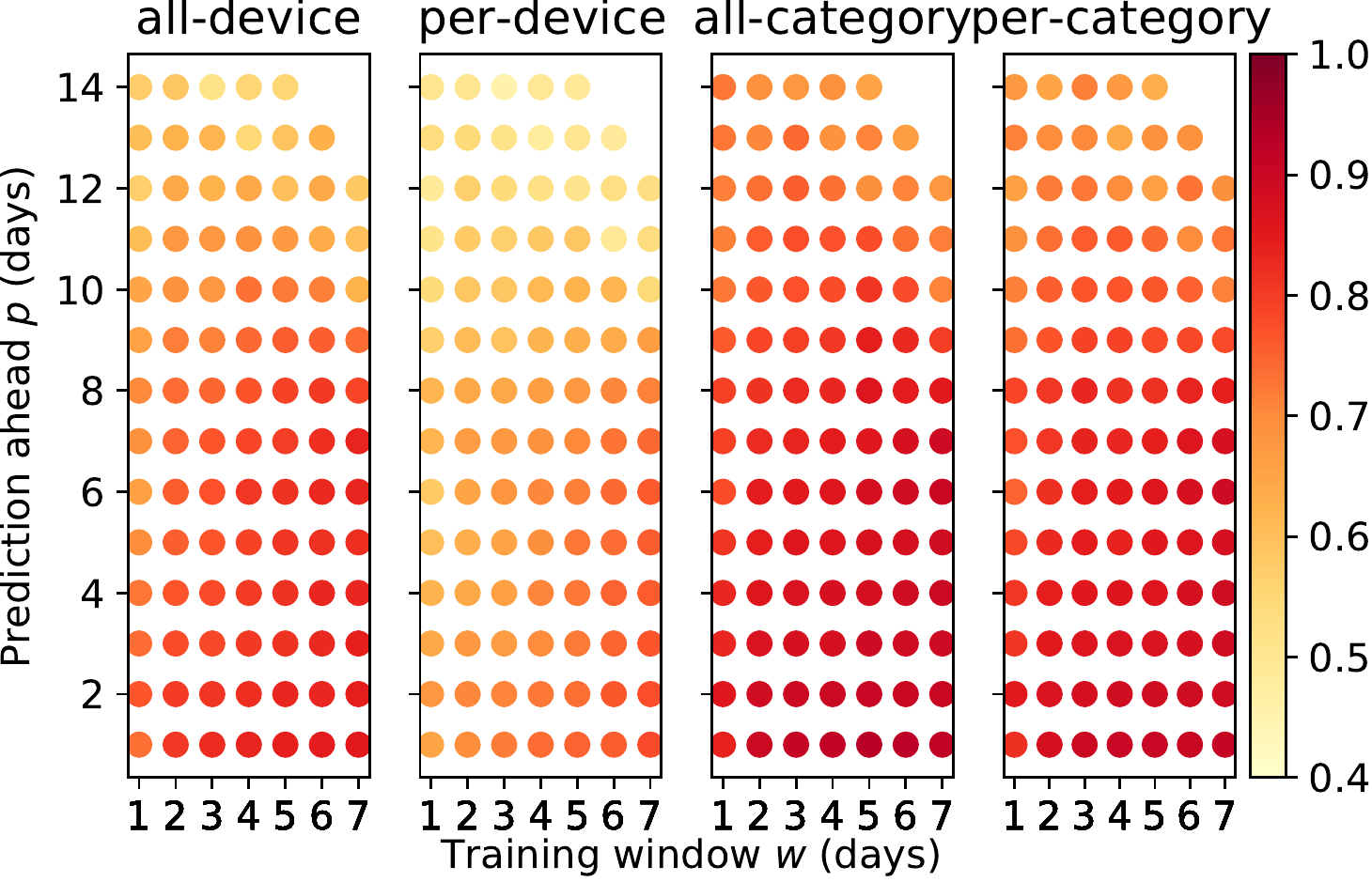}
    \caption{Decision tree classifier}
    \label{fig:testbed_pred_dtc}
    \end{subfigure}

    \caption{\fscore of various classifiers. The x-axis shows the length of the
    training window $w$, the y-axis shows the prediction day $p$. The color at
    the given $(w \times p)$ coordinate corresponds to the \fscore value. The
    darker the color, the more precise the classifier is.}
    \label{fig:testbed_pred}
\end{figure*}

The input size for all the models was the number of features \s{traces}, \ie 19.
FC, LSTM and Conv1D layers used Rectified Linear Unit (ReLU) activation
layer. For a single model for all devices or all categories, the Output Layer
was a FC layer with the same number of classes as we tried to classify, \ie
43 and 6 for \emph{device} and \emph{category} respectively, with SoftMax
activation and categorical cross-entropy loss function. For a single model
per device or category, the Output Layer was a single neuron with Sigmoid
activation and binary cross-entropy loss function. 

In the case of neural networks we trained the models for 5 epochs with batch
size = 128. We have tried to train the models for the larger number of epochs, but
it rarely led to an increase in accuracy by more than 1 percentage  and the
time spent on training was significantly higher.

In this paper our work focuses on on-line device classification, hence we focus
on models which are simple and lightweight enough to run on an edge device,
rather than the larger optimized models.  Therefore, we use features which can
be extracted almost immediately and we do not rely on statistical data from a
long period of time (\eg number of IP addresses contacted over an hour or a
number of DNS requests).

%% file: classification.tex
\section{Device/Category Classification}
\label{s:classification}

We used a standard evaluation metric~\cite{Hackeling2014}, \fscore for the overall measure of the  models accuracy and is defined as:

\[ F_1 = 2 \times \frac{precision \times recall}{precision + recall}, F_1 \in
\langle 0, 1 \rangle \]

It represents a harmonic mean of precision and recall. 

\subsection{Training Window vs. Prediction} We evaluated the \fscore of more
than 42,000 machine learning models and analyzed how the length of the training
window size $w$ influence the \fscore of classification of the network flow of
prediction day $p$.  Results for each model type can be seen in
\fig{testbed_pred}. Each figure is split into four separate sub-figures, each
depicting results for a different model group (\emph{all} vs. \emph{per} and
\emph{device} vs. \emph{category}). 

The x-axis
shows the length of the training window $w \in \langle 1 \ldots 7 \rangle$,
while the y-axis shows the prediction day $p$. The prediction day is capped at
$p = 14$, \ie we evaluate prediction for up to 2 weeks ahead. The darkness of
the dot at the corresponding coordinate shows the \fscore for a given training
length window $w$ and the prediction day $p$. We are omitting three data points
in the top right corner, \ie $p = {13, 14}$ where $w = 7$ and $p = 14$ where $w
= 6$ due to the lack of data-points to calculate an average. Because the number
of days in our dataset $d_{max} = 21$, in the case of $w = 7$, we obtain a
single data point for $p = 14$ which we do not consider statistically confident.

Generally, three trends can be observed: \one the color gets lighter (\ie
\fscore decreases) with the decrease of the size of the training window $w$
(from right to left), \two the color gets lighter with the increase of the
number of days of prediction $p$ (from bottom to top), and \three for
predictions close to two weeks ahead the length of the window does not have an
impact on the accuracy of the prediction (the color stays virtually the same for
$p \geq 12$). 

Analysis of the results shows that the largest difference in \fscore is when $w$
increases from 1 to 2. Increasing the training window from one day to two
increases the \fscore on average by almost $0.05$. Further increase of the
training window usually leads to higher \fscore but the increase is not so
dramatic. Increasing the training window from six to seven days improves the
\fscore on average by less than $0.01$ 

Generally, it can be observed that the \fscore steadily decreases with the
increase of the prediction day $p$. The rate at which the \fscore decreases is
determined by the length of the training window. It can be expected that the
model trained on a larger dataset will perform better over a longer period of
time.  However, analysis shows that after 7 days, the \fscore decreases more
rapidly. This holds for all types of models. On average, the \fscore decreases
by $0.01$ per day for the first seven days, followed by the average decrease
of $0.023$ per day for the following seven days. This trend is less visible
for the models with shorter training window $w$ and more visible for models with
longer training windows. For predictions for $p = 14$ days, the \fscore oscillates
around $0.5$ for device and $0.6$ for the category classification, but there is no
increasing or decreasing trend depending on the length of the window.

\takeaway{This suggests that none of the models can reliably classify devices
more than two weeks ahead and therefore regular update of the model is
required.}

\subsection{Model Types \& Model Groups}

\begin{figure}
    \centering
\includegraphics[width=0.45\textwidth]{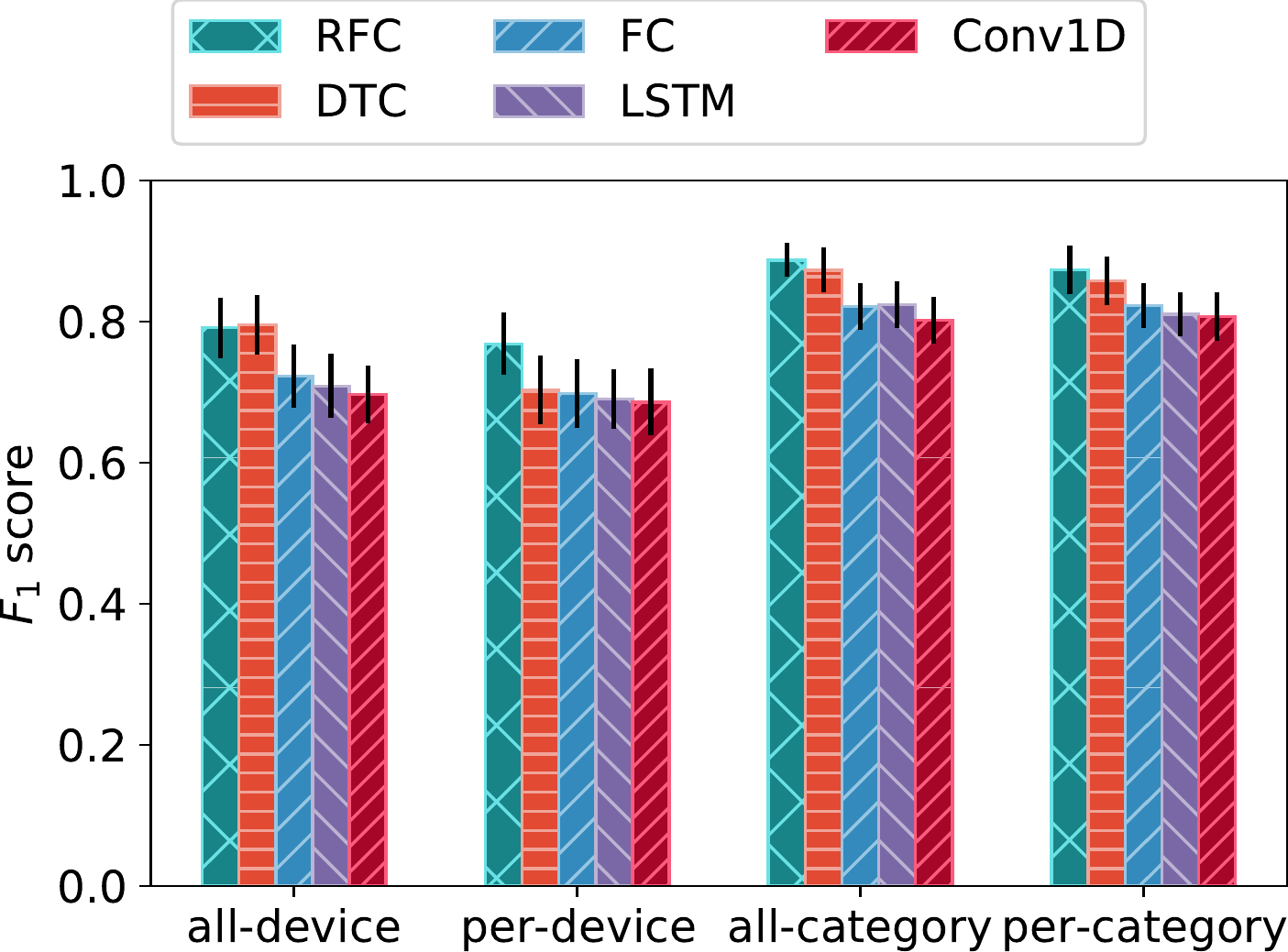}
    \caption{Average \fscore of various models using a training window of
    various sizes (1-7) over the prediction of up to 7 days ahead.}
    \label{fig:prediction}
\end{figure}

We compare the \fscore achieved by different model types (RFC, DTC, FC, LSTM,
Conv1D) and model groups (\emph{all} vs. \emph{per} and \emph{device} vs.
\emph{category}). For each model and each training window we average the \fscore
of predictions for one week ahead.  \fig{prediction} shows the
average of averaged \fscore grouped by model group and model type.

\begin{figure*}
    \centering
    \begin{subfigure}[t]{.45\linewidth}
\includegraphics[width=\linewidth]{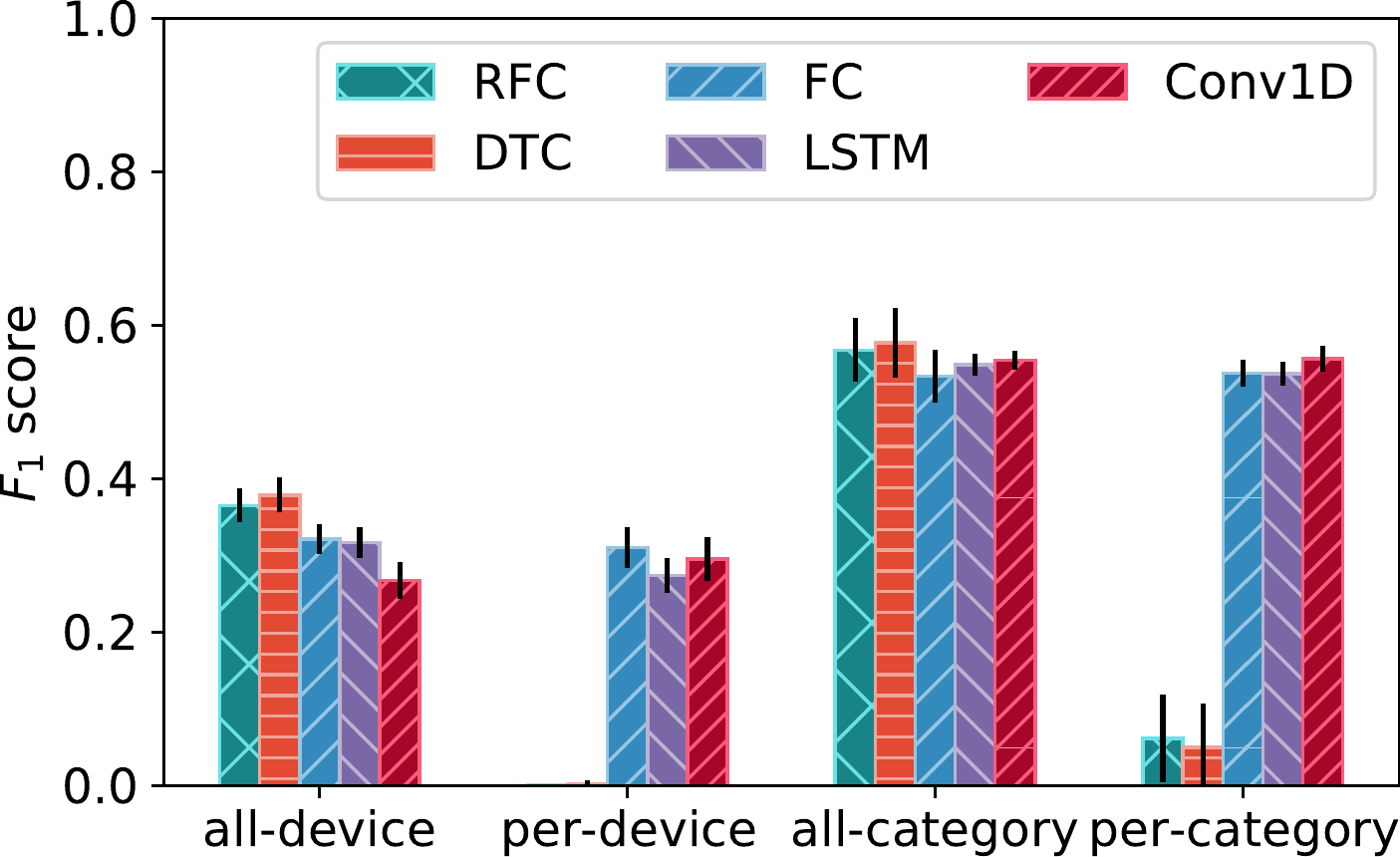}
    \caption{Large Test-bed}
    \label{fig:active_testbed}
    \end{subfigure}
    \begin{subfigure}[t]{.45\linewidth}
\includegraphics[width=\linewidth]{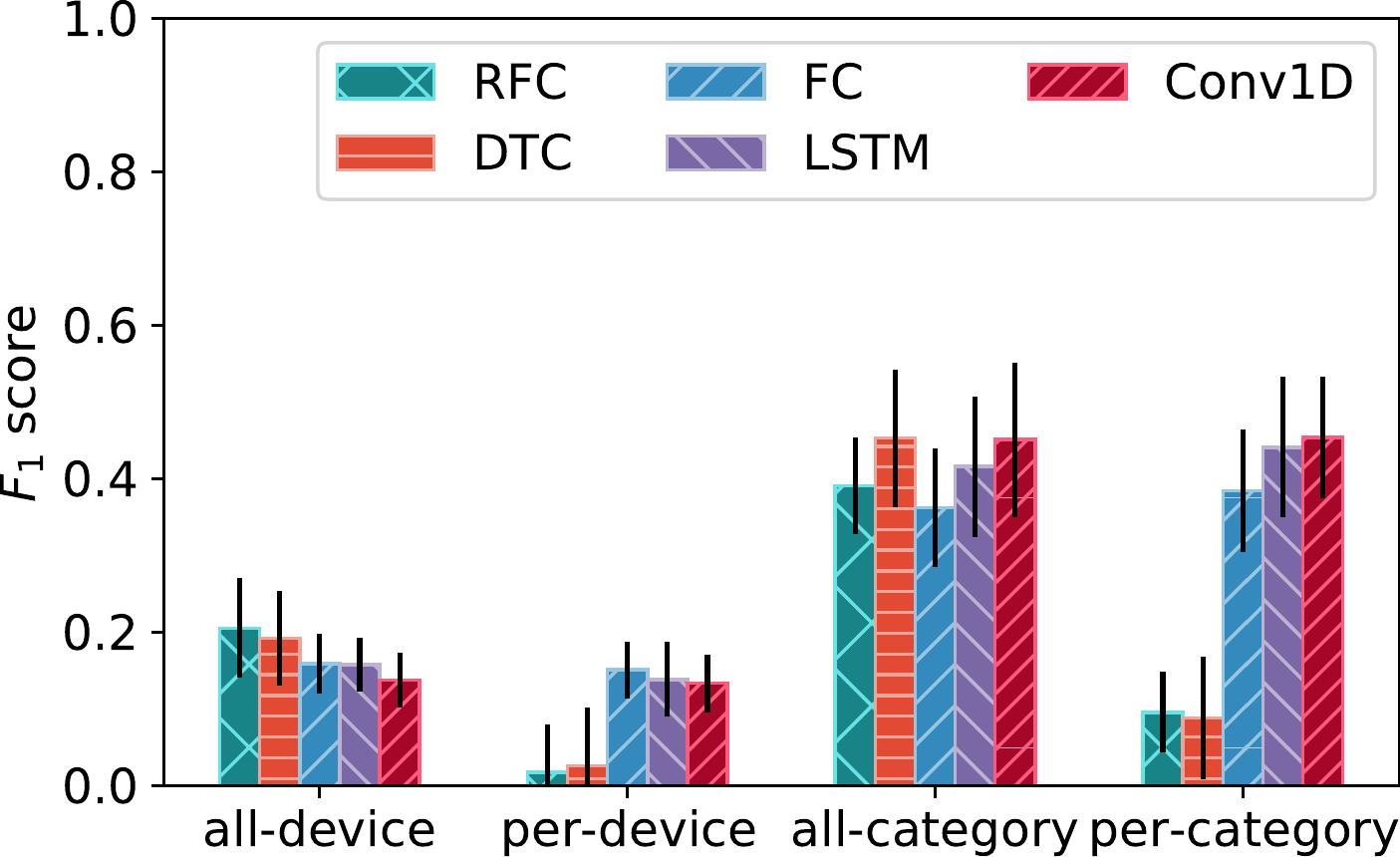}
    \caption{Small Test-bed}
    \label{fig:active_pi}
    \end{subfigure}
    \caption{Average \fscore of models trained on 7 day window of idle data and
    tested on active data of the large and the small test-bed.}
    \label{fig:idle_on_active}
\end{figure*}

The RFC and DTC model perform very similarly in all but the \emph{per-device}
model group. In all cases they slightly outperform the models based on neural
networks. The average difference of \fscore between the tree-based and the NN
models is $0.06$

Generally speaking, all models perform better with the category classification
than with the device classification. The average \fscore for the category
classification is $0.84$ and for the device classification is $0.73$ 

Multiclass classifiers overall slightly outperform multiple binary classifiers
on average by $0.02$. However, the difference is slightly more visible in the
case of device classification where the difference in \fscore is $0.034$
compared to the category classification where the difference is $0.007$. 

\takeaway{RFC and DTC models marginally outperform all models based on neural
networks. The performance of neural network based models is essentially the
same. Creating a single multi-classification model slightly outperforms an
approach based on multiple binary classification models. This applies to both
device and category classification.}

\begin{figure*}
    \centering
    \begin{subfigure}[t]{.33\linewidth}
\includegraphics[width=\linewidth]{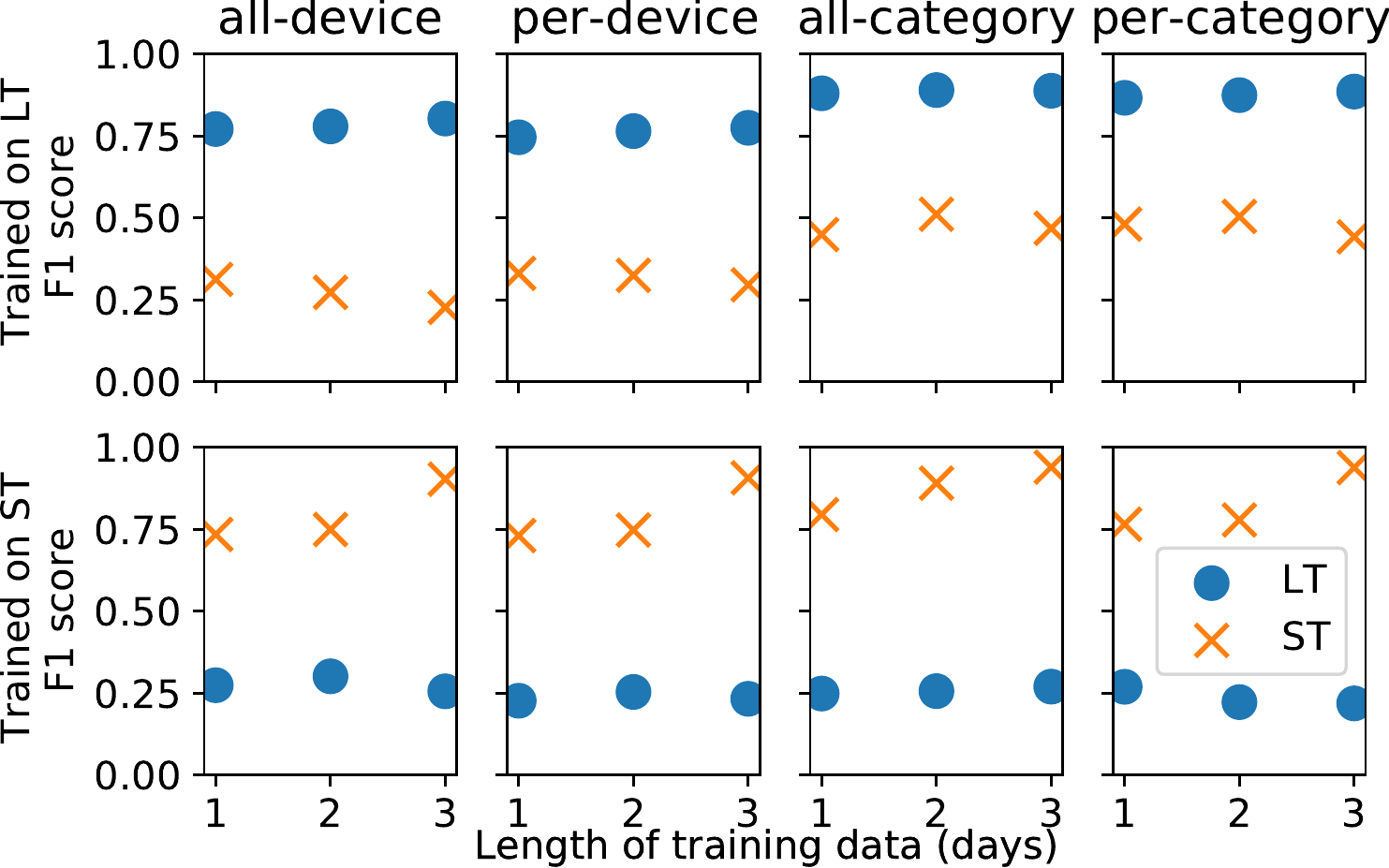}
    \caption{Fully connected model}
    \label{fig:testbed_nn}
    \end{subfigure}
    \begin{subfigure}[t]{.33\linewidth}
\includegraphics[width=\linewidth]{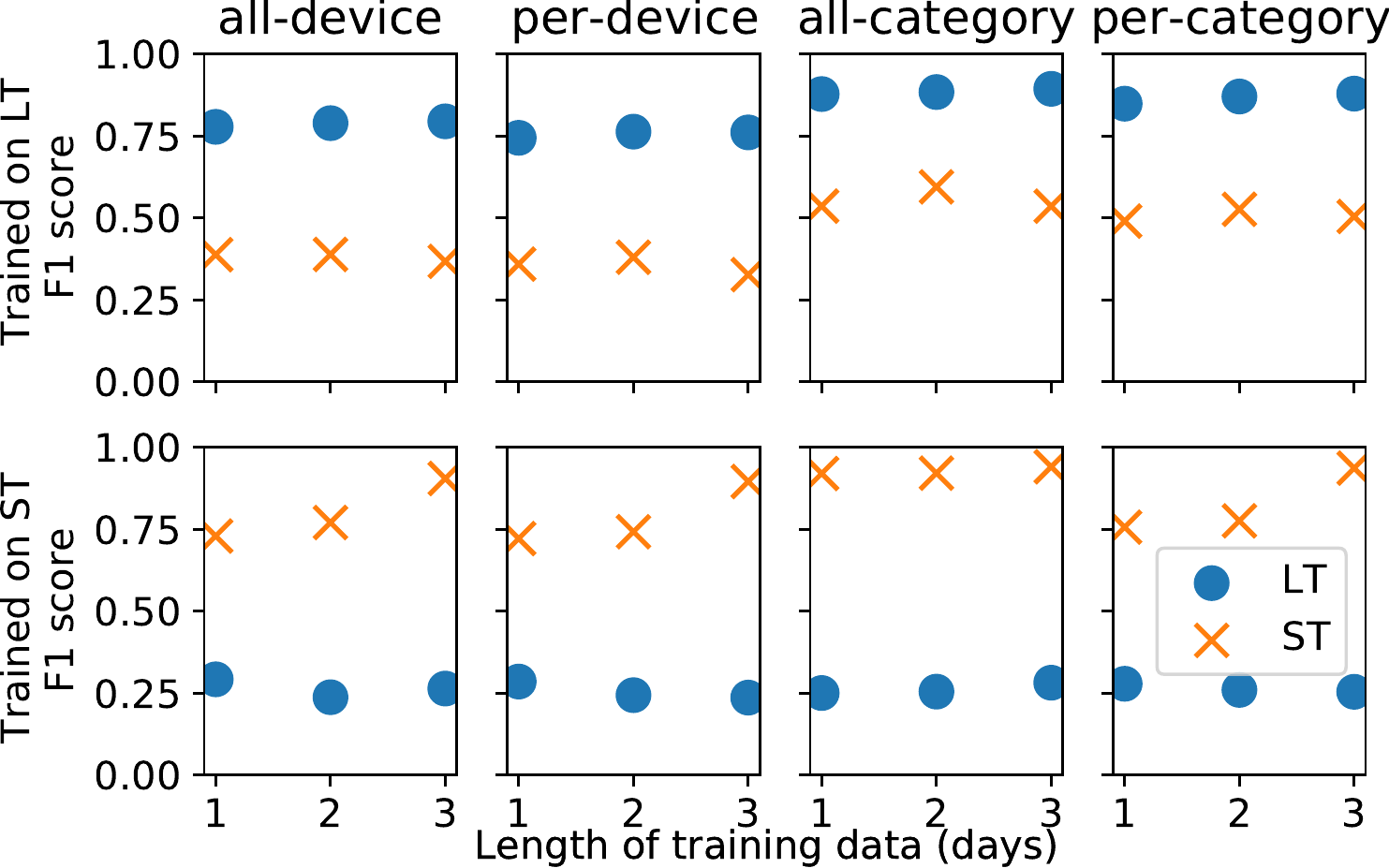}
    \caption{LSTM model}
    \label{fig:testbed_lstm}
    \end{subfigure}
    \begin{subfigure}[t]{.33\linewidth}
\includegraphics[width=\linewidth]{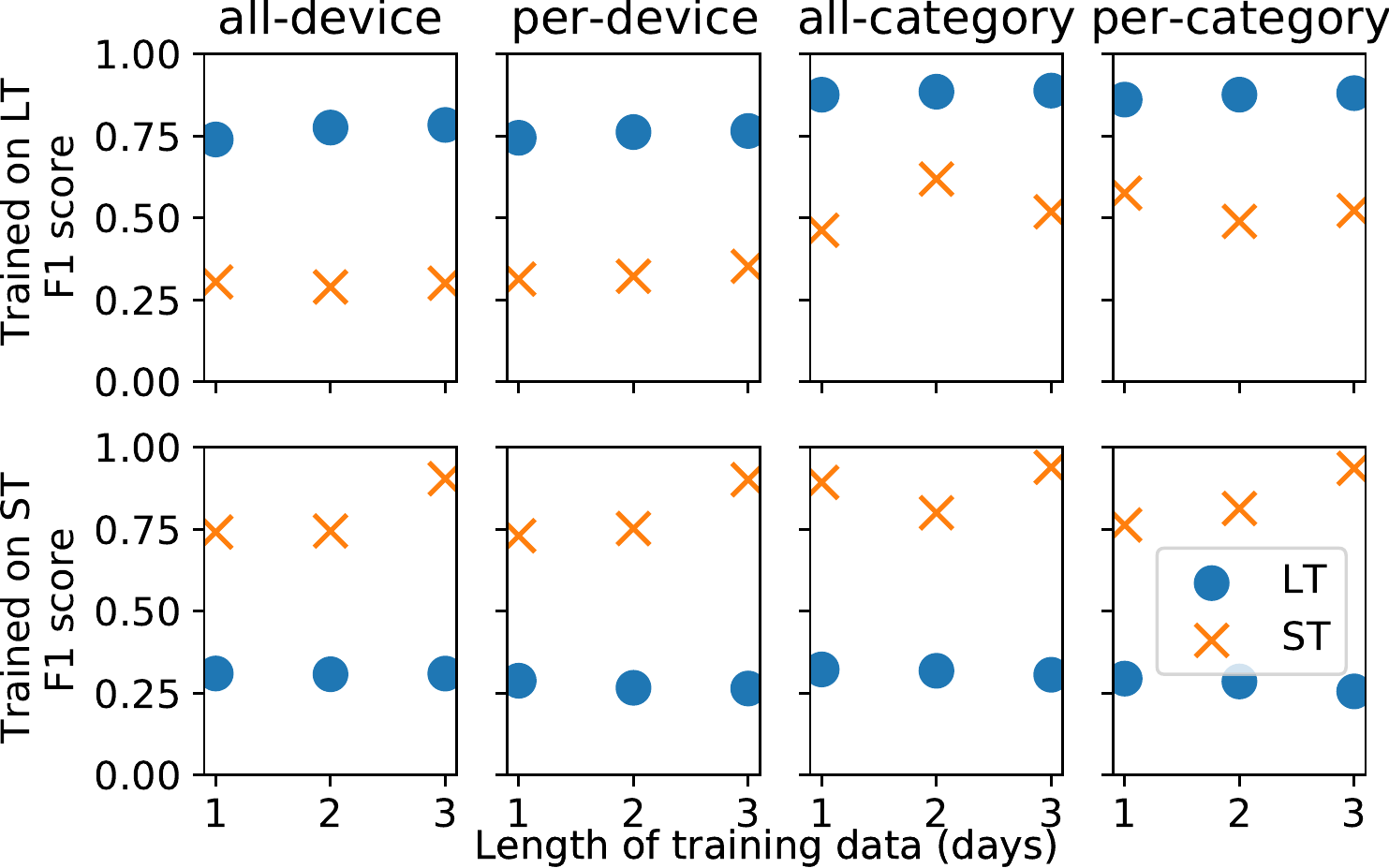}
    \caption{1D convolutional model}
    \label{fig:testbed_conv1d}
    \end{subfigure}
    \caption{Comparison of \fscore of base models (trained on idle data) and
    updated on the large test-bed (LT, top row) and the small test-bed (ST, bottom
    row). The mark shows the average \fscore for device/category classification
    achieved over 7 days of active experiments. The x-axis shows how many days
    (1-3) of data were used to update the model. \textopenbullet~denotes score
    achieved on LT, while \texttimes~denotes score achieved on ST.}
    \label{fig:testbed_eval}
\end{figure*}

\subsection{Active Dataset}

Models trained on the \emph{idle} dataset show a reasonably high \fscore
suggesting that they can be used for on-line device/category classification of
\iot devices in a home environment. Next, we evaluate these models on data
collected over seven days of automated experiments on our two test-beds
\s{dataset}: \one the \emph{large test-bed (LT)} and \two the \emph{small
test-bed (ST)}. The \emph{active} data consist of the mixture of
light/medium/heavy usage patterns as described previously \s{dataset}.  For the
purpose of evaluation we chose the models which achieved the highest \fscore on
the idle data from the large test-bed. These models were trained on a seven day
window $w = 7$. We evaluated these models on active data from both the large and
small test-bed. For the purpose of fair evaluation, we included only the devices
that were in common in both test-beds.

\fig{idle_on_active} shows the comparison of all five models on the large
(\fig{active_testbed}) an the small (\fig{active_pi}) test-bed. Two facts can be
observed: \one models trained on the idle data of the large test-bed achieve
higher \fscore on active data collected from the same test-bed rather than the
small test-bed, despite containing the same type of devices, \two even if the
models are trained on the data from the large test-bed, the device/category
classification is significantly worse when compared to classification of idle
data. 

The \fscore achieved by all three models based on neural networks was rather
similar. The average score achieved on the large test-bed for device (category)
classification was $.3$ ($0.54$, respectively). This score is significantly
lower than the score achieved on idle data. However, when the same models were
tested in the small test-bed, the \fscore was halved to only $0.15$ for device
classification and decreased to only $0.42$ for the category classification. 

RFC and DTC models achieved slightly higher \fscore when compared to models
based on neural networks, but only in the case of a single model for \emph{all}
devices/categories. However, this score is still significantly lower than the
score achieved on the idle dataset. Surprisingly, the \emph{per-*} model group
performed extremely badly when evaluated on the active data. We believe
individual models were very fine-tuned for specific type of traffic and
therefore achieved very low \fscore on a different type of traffic.

This fact supports our argument that it is necessary to keep the models updated
locally, with data collected from the household. The results also show that
models updated in one household are not applicable in another household, even if
the devices are the same.

Because RFC and DTC models do not support updating of the models with new data,
we omit them from the further evaluation. As the models trained on the idle
dataset performed so poorly on both of active datasets, we updated the models
with the data collected during the active experiments. First, we chose the model
that achieved on average the highest \fscore on all dates of both test-beds.
Unsurprisingly, it was a model trained on a 7 day window of data. We refer to
this model as the \emph{base model}. We also tested a model trained on the
\emph{whole} idle dataset. Surprisingly, this model achieved slightly lower
average \fscore than the best model trained on a 7 day window.  This fact
suggests overfitting of the model might be a problem and a smaller dataset might
achieve higher \fscore.

Next, we update the \emph{base model} with one, two, or three days of data
either from the large or the small test-bed. \fig{testbed_eval} depicts the
comparison between these updated models on both test-beds. Two important facts can be
observed immediately: \one updating the base model with active data from the test-bed
significantly increases the \fscore of device/category classification on the
\emph{same} test-bed and \two updating the base model with data from one
test-bed increases the \fscore of device/category classification of the
\emph{other} test-bed only marginally. These two facts can be observed for both
test-beds. 

Updating the \emph{base model} with just one day of active data from the large
test-bed increases the average \fscore from just $0.3$ to $0.75$ for the device
classification and from $0.54$ to $0.87$ for the category classification. Each
additional day increases the \fscore on average by $0.011$. On
the other hand, when the same model is evaluated on the small test-bed, the
\fscore increases only from $0.15$ to $0.33$ for the device classification and
from $0.42$ to $0.5$ for the category classification. Increasing the training
dataset by additional days does not yield a better \fscore. 

Similarly, when the \emph{base model} is updated with only one day worth of data
from the small test-bed, the average \fscore raises from just $0.15$ to $0.73$
for the device classification and from just $0.42$ to $0.82$ for the category
classification.  Adding one more day worth of data increases the \fscore by
additional $0.02$ in the case of device and $0.015$ in the case of
category classification.  Using data from all three days, a \fscore as high
as $0.9$ (up from just $0.15$) for device and $0.94$ (up from just $0.42$) for
category classification is achieved.  On the other hand, updating the model with
the data from the small test-bed has very limited impact on the \fscore of the
large test-bed classification. The average \fscore for the device and the
category classification is only $0.27$, and it remains rather stable regardless of
the number of days used to update the model. 

\takeaway{Models trained on a dataset collected from an idle test-bed are not
accurate and fail when are used on active data. Even a small amount of active
data can significantly increase the accuracy of the models. However, a model
retrained with active data from one test-bed is not applicable on the same type
of data in another test-bed. It means that in order to achieve high accuracy,
the model needs to be updated with local data.}

\subsection{Model sizes}
\label{s:model-sizes}

The model size is an important factor for being able to run inference on an edge
device, because such a device usually has limited main memory. During our first
attempts, the RFC and DTC models did not fit into the main memory, hence
inference could not be run at the edge device. Therefore, for these two cases,
we performed a hyperparameter (\eg the number of estimators or the depth of the
tree) search to tune the model size while maintaining comparable accuracy. 

\begin{figure}
    \centering
    \includegraphics[width=0.45\textwidth]{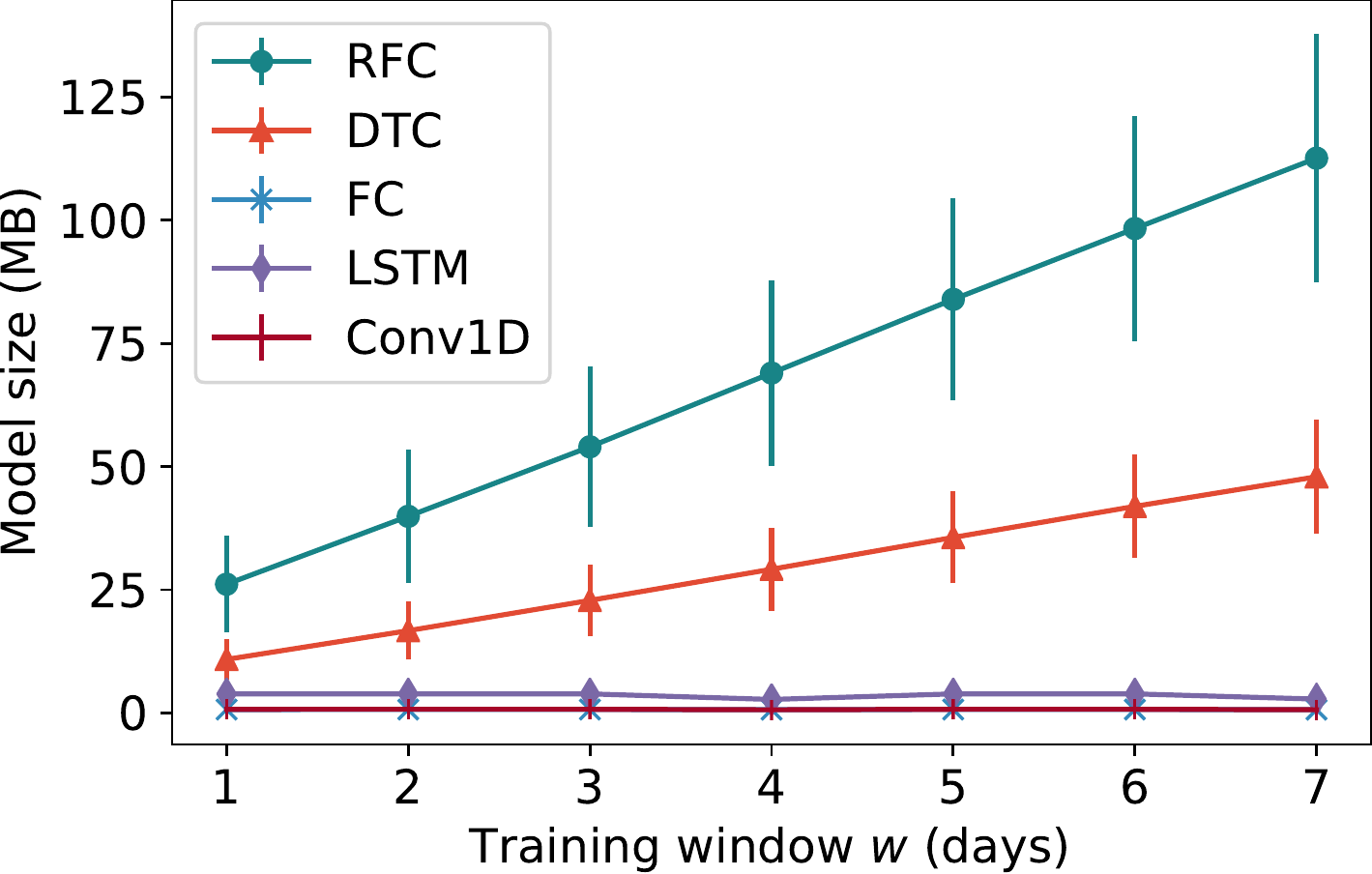}
    \caption{Average model sizes for different types of models for various
    training windows. The models are from the \emph{all-device} group.}
    \label{fig:model-sizes}
\end{figure}

\begin{figure*}[!ht]
    \centering
    \begin{subfigure}[t]{.33\linewidth}
\includegraphics[width=\linewidth]{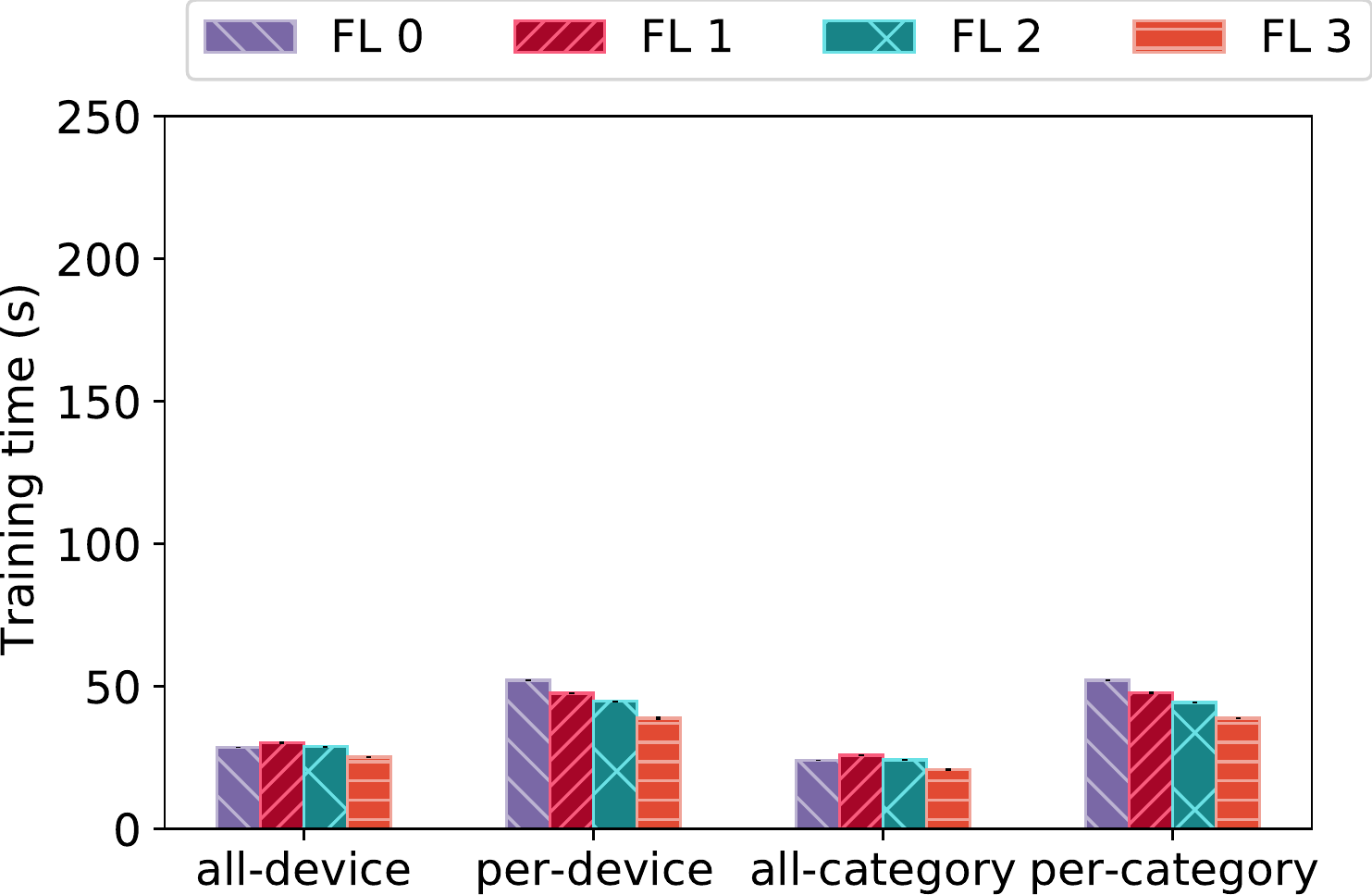}
    \caption{Fully connected model}
    \label{fig:training_freeze_nn}
    \end{subfigure}
    \begin{subfigure}[t]{.33\linewidth}
\includegraphics[width=\linewidth]{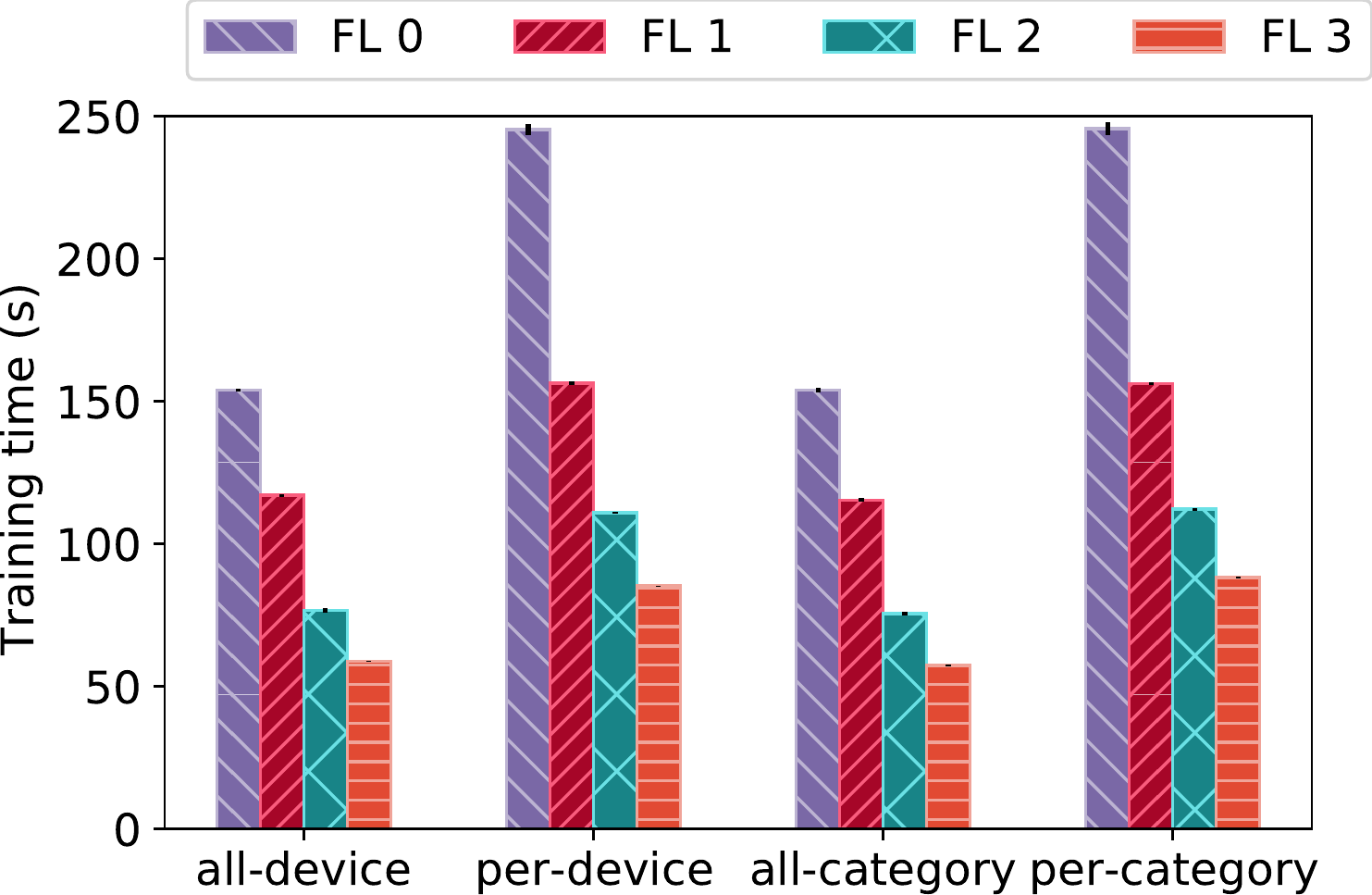}
    \caption{LSTM model}
    \label{fig:training_freeze_lstm}
    \end{subfigure}
    \begin{subfigure}[t]{.33\linewidth}
\includegraphics[width=\linewidth]{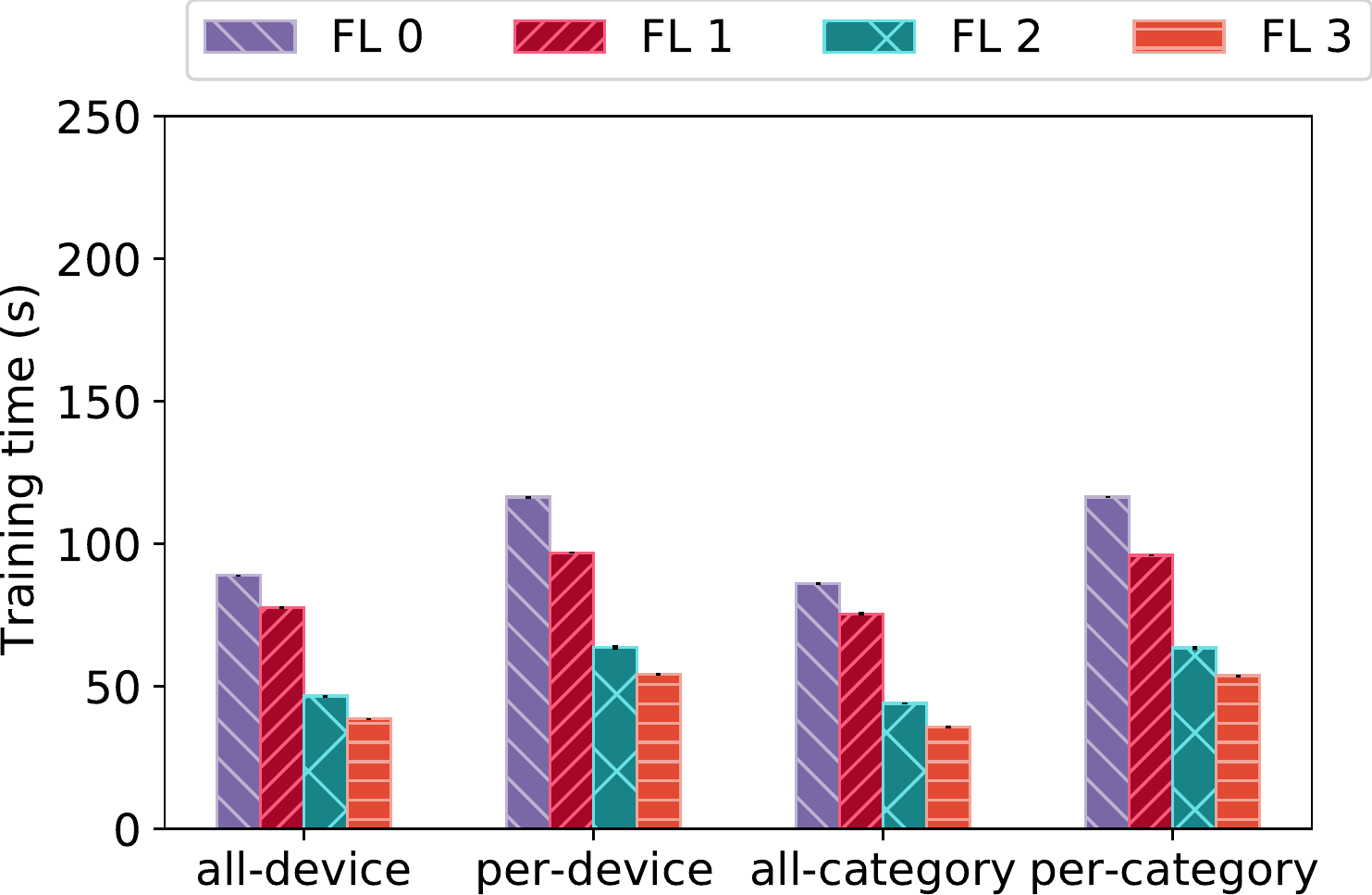}
    \caption{1D convolutional model}
    \label{fig:training_freeze_conv1d}
    \end{subfigure}
    \caption{Average training time on RPi4 with different numbers of frozen layers (0, 1, 2 or 3 layers).}
    \label{fig:training_freeze}
\end{figure*}

\fig{model-sizes} shows the sizes of the models in MB for all types of ML models
depending on the length of the training window $w$. The figure shows the average
size of the \emph{all-device} group of models.  The model sizes of RFC and DTC
are increasing linearly with the length of the training window, as the size
depends on the size of the training set. On the other hand, the size of the
models based on neural networks remains constant. 
In this case, the type, size, and number of layers influence how many weights
there are in the model, and this influences their final size.

\takeaway{The model size of RFC and DTC depends on the number of data points in
the training dataset. As we have shown previously, the models need to be
updated, and they need to updated locally. Therefore, solutions based on RFC or
DTC are not scalable. On the other hand, models based on neural networks can be
updated without training from scratch, and also their size remains constant.
Therefore they are suitable for being deployed at the edge.}

%% file: edge.tex
\section{Training \& Inference at the Edge}
\label{s:edge}

Since model accuracy decays over time, the models need to be updated frequently
through retraining using new incoming data in order to maintain good accuracy.
But retraining models in a centralized manner requires shipping data from home
\iot networks to the cloud, which presents challenges in terms of scalability
and, more importantly, privacy preservation. Instead, updating the model can
take place at the edge of the network, on the home gateway device.  This has the
advantage of not only reducing the communication bandwidth and latency between
the gateway device and the central server, but also preserving the user privacy. 
On the other hand, edge devices have constraints in terms of CPU and
memory. In this section, we examine the impact of these constraints on the
retraining of models explored in previous sections, and on running inference on
an edge device. To reduce the computation overhead associated with retraining of
the model, we reuse parts of the globally trained NN model by freezing layers of
the model. Furthermore, we show that our models can run on a representative edge
device, demonstrating that \iot devices can be identified at the edge.

As a representative edge device, we used Raspberry Pi model 4 with
4~GB of RAM. The Raspberry Pi is running the Ubuntu 19.04 operating system with
Python 3.6, on which a traditional ML stack was installed (numpy, scipy, pandas,
scikit-learn, tensorflow, and keras packages).

\subsection{Speed of Model Retraining at the Edge}

Given the fact that the \fscore of the models decreases over time, we
investigated the possibility of updating the model locally at the edge using new
traffic data from that respective home \iot network. However, we did not start
training from scratch, but we used the global model downloaded from a central
server. Additionally, we evaluated how freezing different numbers of layers of
the neural network (zero, one, two and three layers) influences the training
time and \fscore of the model. Freezing zero layers means updating the whole
model, \ie  re-computing all weights, while freezing three layers means updating
only the last layer, \ie re-computing only a small fraction of weights. We use the
idle dataset in our retraining experiments.
\fig{training_freeze} presents the average running time for training the neural
network models on the selected edge device (Raspberry Pi 4). In the case of the
\emph{per-device} models, we plot the training time for training a single model
for a single device. We trained models for all of our devices, and the training
time for each device is approximately the same. Training models for
multiple devices on a Raspberry Pi in our testbed takes
$43 \times t_{per-device}$, where $t_{per-device}$ is the average training
time for a single device. Similarly, we plot the training time for training a
single model for a single category. As such, the total training time for our six
categories is $6 \times t_{per-category}$, where $t_{per-category}$ is the
average training time for a single category.

For the FC models, in the case of \emph{all-device} and \emph{all-category}
models, freezing one or two layers does not reduce the training time, while
freezing three layers reduces the training time by $11.8\%$ and by $13.3\%$
respectively. This is probably caused by the structure of the FC model, where
the number of neurons increases in each layer, \ie most weights are in the last
three layers. The average training time with three frozen layers is $25$~s for
\emph{all-device}, and $20$~s for \emph{all-category}. On the other hand, in the
case of LSTM and Conv1D models, the training time decreases with each additional
frozen layer. For LSTM in the case of \emph{all-device} models, one frozen layer
lowers the training time by $24\%$, two frozen layers by $50.2\%$, and three
frozen layers by $61.7\%$.  The average training time using three frozen layers
is $59$~s.  For Conv1D \emph{all-device} models, one frozen layer reduces the
training time by $12.7\%$, two frozen layers by $47.6\%$, and the three frozen
layers by $56.6\%$.  The training time with three frozen layers is $38$~s.  The
results for \emph{all-category} are just slightly smaller compared to
\emph{all-device} for both LSTM and Conv1D.

The training times are approximately the same for the \emph{per-device} and
\emph{per-category} models for all three model types. For brevity we discuss
only the \emph{per-device} results.  In the case of the FC models, freezing one
layer reduces the training time by $8.5\%$, two layers by $14.1\%$, and three
layers by $25.6\%$. The training time when freezing three layers is $38$~s.  In
the case of LSTM models, freezing one layer reduces the training time by
$36.2\%$, two layers by $54.8\%$, and three layers by $65.2\%$. The average
training time when freezing three layers is $85$~s.  Lastly, for Conv1D models,
freezing one layer reduces the training time by $16.7\%$, two layers by
$45.3\%$, and three layers by $53.4\%$. The average training time using three
frozen layers is $54$~s.

\takeaway{Model retraining is feasible at the edge for a common household for
all types of models. The improvement in training time largely depends on the
type and the architecture of the neural network.  Layer freezing more than
halves the training time for LSTM and Conv1D models, while for FC layer freezing
reduces the training time modestly.}




\begin{figure*}
    \centering
    \begin{subfigure}[t]{.33\linewidth}
\includegraphics[width=\linewidth]{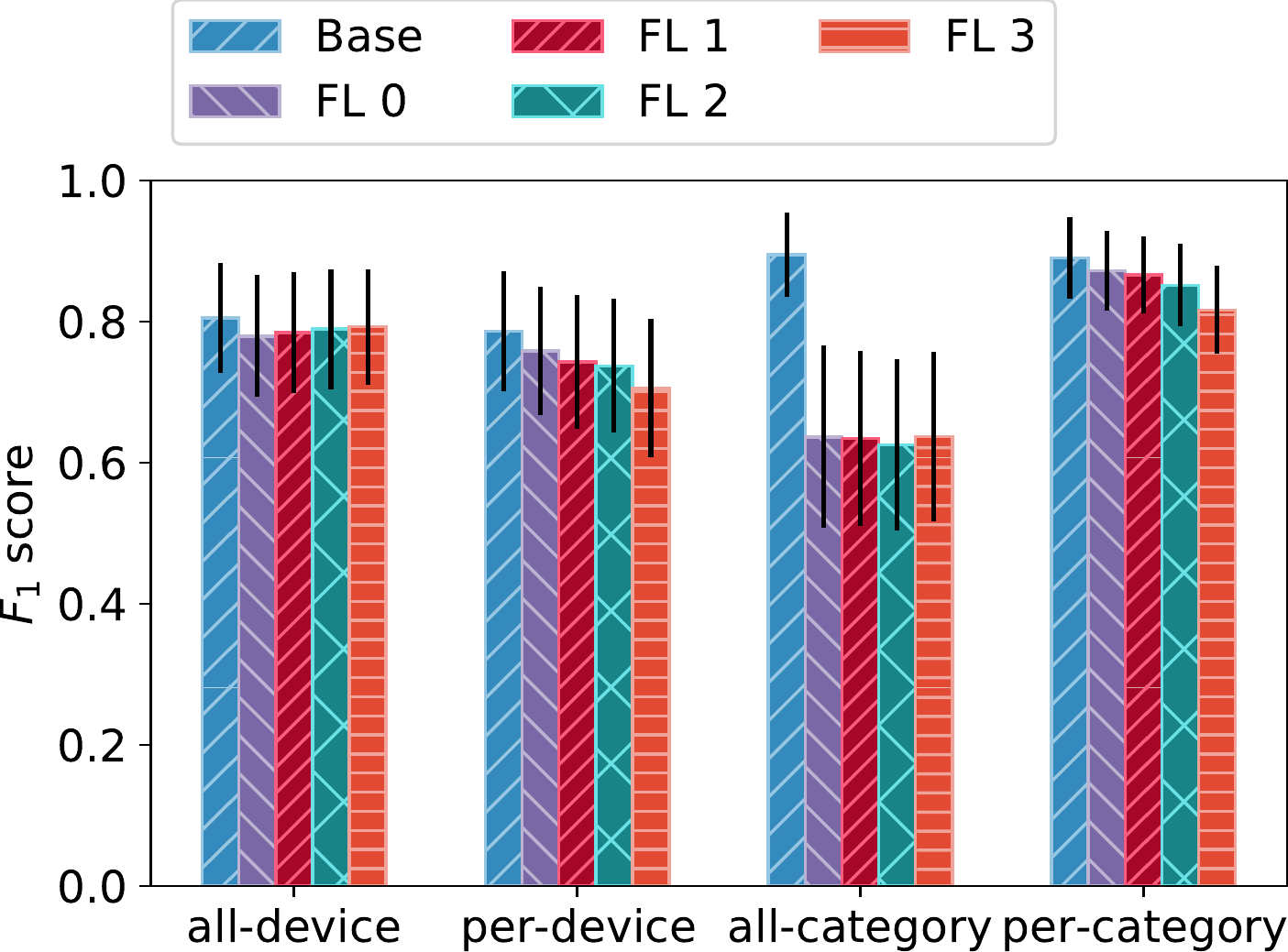}
    \caption{Fully connected model}
    \label{fig:pred_freeze_nn}
    \end{subfigure}
    \begin{subfigure}[t]{.33\linewidth}
\includegraphics[width=\linewidth]{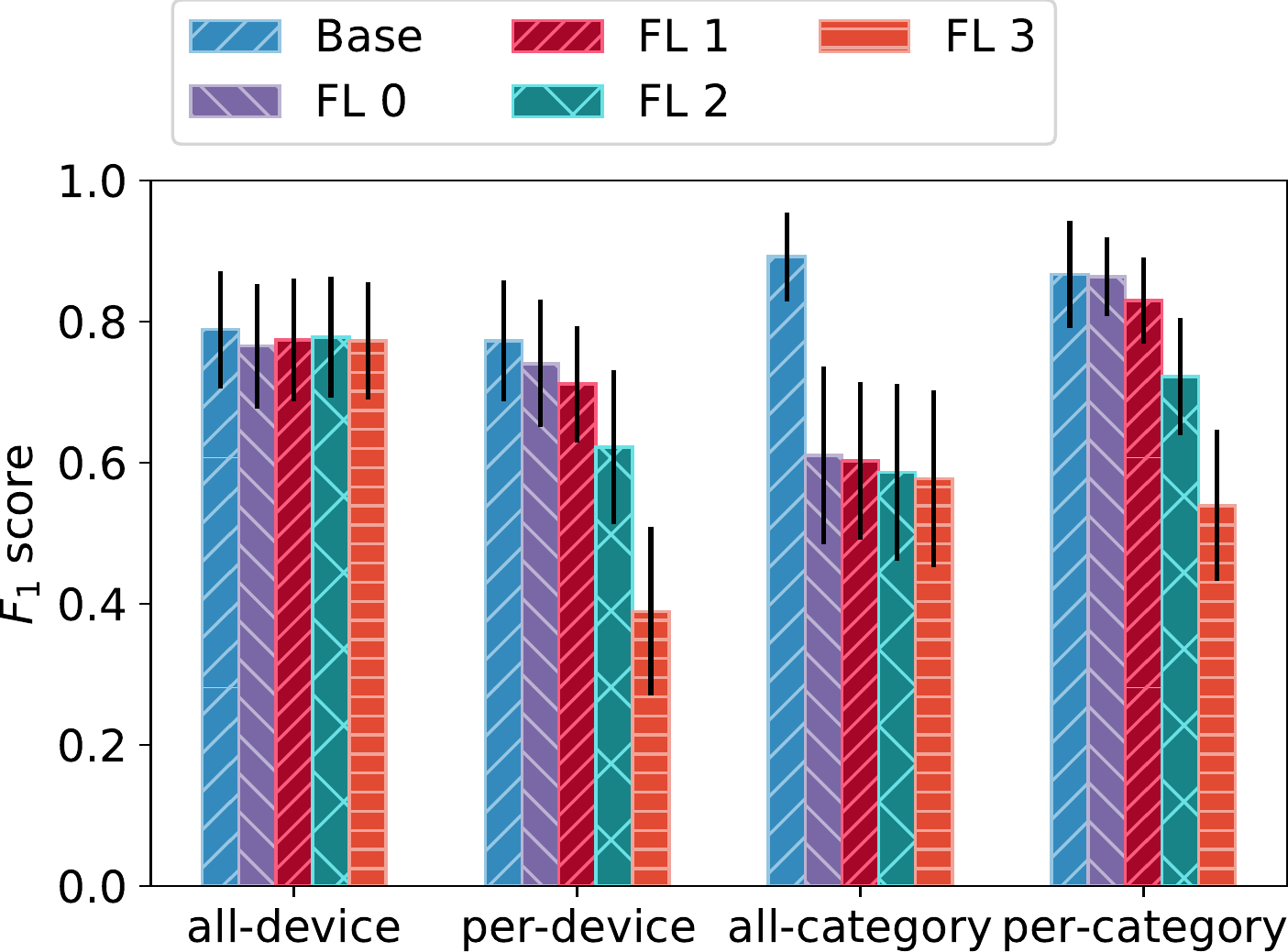}
    \caption{LSTM model}
    \label{fig:pred_freeze_lstm}
    \end{subfigure}
    \begin{subfigure}[t]{.33\linewidth}
\includegraphics[width=\linewidth]{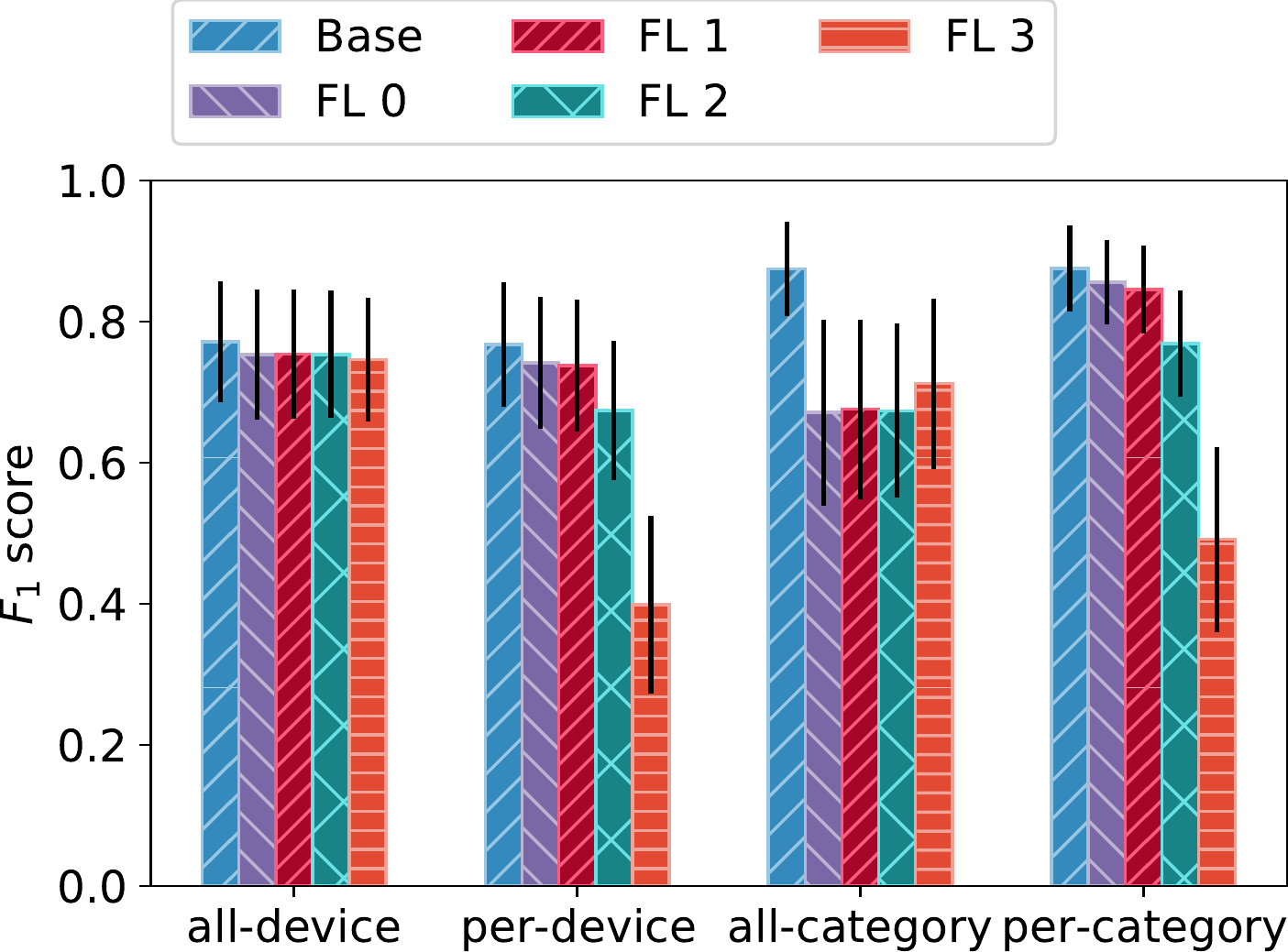}
    \caption{1D convolutional model}
    \label{fig:pred_freeze_conv1d}
    \end{subfigure}
    \caption{Comparison of \fscore with various number of frozen layers.}
    \label{fig:pred_freeze}
\end{figure*}

\subsection{$F_1$ Score of Retrained Models}

To demonstrate how freezing of layers impacts model's \fscore we selected all
models already trained on a four day window. The model was updated with the
following three days of data while either not freezing any layer or freezing one
to three layers.  These updated models are effectively trained on seven days
worth of data and therefore are compared with models trained on a seven day
window. We refer to this model as the \emph{base model}. 

\fig{pred_freeze} shows how freezing of different numbers of layers influences
different types and groups of neural networks. In the case of a
\emph{all-device} classification, the \fscore is on average between $0.016$ and
$0.02$ lower for all types of neural networks. Surprisingly, the
\fscore remains virtually the same regardless of the number of frozen layers. 

More surprising is the case of the \emph{all-category} classification. In this
case the \fscore drops by $0.25$ in the case of FC network, $0.28$ in
the case of LSTM, and $0.2$ in the case of Conv1D. Again, the number of
frozen layers has a very limited impact on the \fscore of the models. The
reason for such a dramatic decrease is yet unknown and would require deeper
inspection and analysis of the weight updates in the model, which is out of the
scope of this paper. 

While in the case of \emph{all-*} group of models there was very little
difference in \fscore when different numbers of layers were frozen, in the case of
\emph{per-*} group of models, the number of frozen layers has significant
influence on the \fscore. The least visible impact is with the FC network, where
freezing an additional layer decreases the \fscore on average by $0.02$ in the
case of \emph{per-device} classification and $0.018$ in the case of
\emph{per-category} classification.

In the case of LSTM models, freezing layers has a much bigger influence on the
\fscore. Freezing between zero and three layers in the \emph{per-device} group,
decreases the \fscore by $0.03, 0.06, 0.15$ and as much as $0.38$ when compared
with the \emph{base model}. Similarly, in the case of \emph{per-category} group,
freezing between zero and three layers leads to decrease of $0, 0.4, 0.14,
\text{and } 0.33$. 

Similarly, in the case of Conv1D model, the influence of freezing different numbers
of layers has a significant impact on the \fscore. Not freezing any layer
decreases the \fscore by $0.02$ points in \emph{per-device} and \emph{per-category}
groups when compared to the base scenario. Freezing one layer reduces \fscore
by $0.03$ (in both scenarios), two layers by $0.09$ and $0.1$ in
\emph{per-device} and \emph{per-category} group respectively, and three layers by as
much as $0.37$ and $0.38$  points. 

\takeaway{ Freezing layers can noticeably decrease the accuracy of the model.
The decrease depends on the type and the structure of the model. In a real
world scenario, the number of frozen layers could either be chosen by the entity
(\eg manufacturer) providing the base model, or by the edge device depending on
its computational power and the number of models requiring updating.}

\subsection{Speed of Model Inference at the Edge}

We run the model inference on the Raspberry Pi 4 to evaluate average inference
running times. We use a lightweight version of the TensorFlow library called
TensorFlow Lite~\cite{tflite}, specifically designed for edge devices.
All models were converted to a format supported by this library. 
In the case of \emph{per-*} models, the
inference is executed on all models sequentially, and the model with the highest
probability is selected as the output device/category. Therefore, the inference
time depends on the number of devices/categories that we need to classify.
Generally, the time scales linearly with the number of devices/categories. 

\begin{figure}
    \centering
\includegraphics[width=0.45\textwidth]{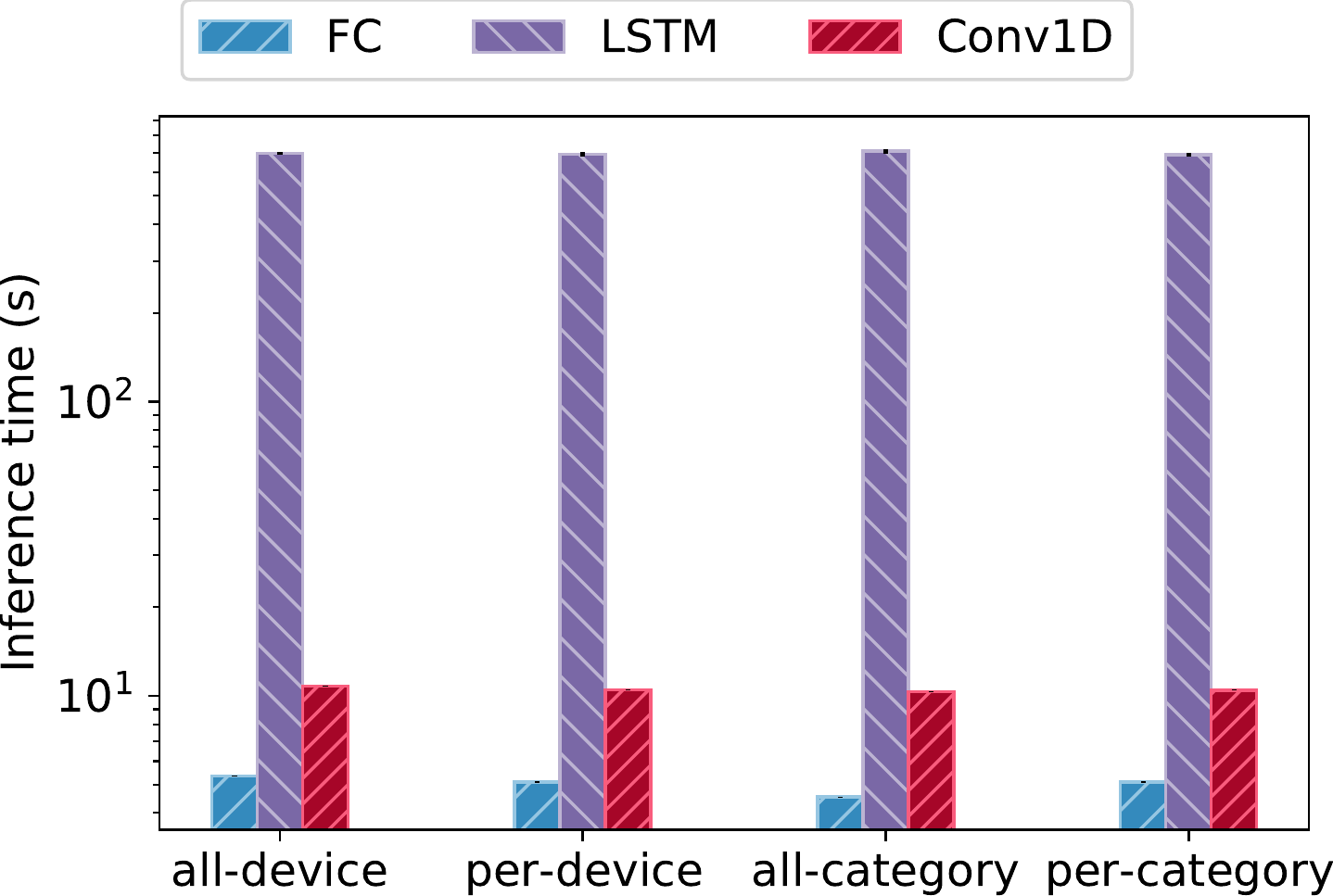}
    \caption{Average inference time on RPi4 of 100K samples using TensorFlow Lite.}
    \label{fig:average-inference-time-100K}
\end{figure}

We run inference using the three neural network models (FC, LSTM, Conv1D), as
detailed in \s{models}.  We first randomly select 1000, 10,000, and 100,000
samples (flows) from our dataset. We feed the samples to each model, and we
measure how long inference takes for this number of samples.
\fig{average-inference-time-100K} presents the results for average inference
time of 100K samples. Results from our experiments show that the total time
scales linearly depending on the number of samples.  Therefore, we omit figures
showing the inference time for 1K and 10K samples.  The average inference time
are very similar for all model groups. We discuss for brevity the results for
the \emph{all-device} group of models.  The average inference time for FC
network is $5.3$~s, for LSTM $697$~s, and for Conv1D is $10.8$~s. 

The results for \emph{per-device} and~\emph{per-category} models are presented
for only one instance of the model being run. Thus, for inference
~\emph{per-device} or~\emph{per-category}, all the models for all of the
devices are run sequentially to determine the device's type or category. This
means that inference takes in our case $43 \times t_{per-device}$,
where $t_{per-device}$ is the average inference time for a single device.
Similarly, the total inference time for our six categories is $6 \times
t_{per-category}$, where $t_{per-category}$ is the average inference time for a
single category.

\takeaway{The results are similar across the four groups of models for each type
of neural network model. However, the inference time for LSTM models is
considerably larger than in the case of FC network and Conv1D models, with FC
being the fastest.  LSTM inference time is approximately $140$ times larger than
FC. Conv1D inference time is double the time of the FC models.}\\

%% file: conclusion.tex
\section{Discussion \& Limitations}
\label{s:discussion}

\begin{center}
\begin{table*}[]
\footnotesize
\captionsetup{skip=0.2em, font=small}
\caption{Classification of an unknown device into a category.}
\label{tab:categorisation}
\begin{tabular}{ l l | l l | l l | l l | l l | l l }
\toprule
\multicolumn{2}{c}{\bf Surveillance} & 
\multicolumn{2}{c}{\bf Media} & 
\multicolumn{2}{c}{\bf Audio} & 
\multicolumn{2}{c}{\bf Hub} & 
\multicolumn{2}{c}{\bf Appliance} & 
\multicolumn{2}{c}{\bf Home automation} \\
\midrule
\bf Model & \bf \fscore &
\bf Model & \bf \fscore &
\bf Model & \bf \fscore &
\bf Model & \bf \fscore &
\bf Model & \bf \fscore &
\bf Model & \bf \fscore \\
\midrule

RFC     &  0.444  &
RFC     &  0.176  &
RFC     &  0.616  &
RFC     &  0.084  &
RFC     &  0.119  &
RFC     &  0.405  \\

DTC     &  0.407  & 
DTC     &  0.168  & 
DTC     &  0.617  & 
DTC     &  0.085  & 
DTC     &  0.120  & 
DTC     &  0.409  \\

FC      &  0.399  &
FC      &  0.154  &
FC      &  0.611  &
FC      &  0.090  &
FC      &  0.078  &
FC      &  0.458  \\

LSTM    &  0.405  &
LSTM    &  0.135  &
LSTM    &  0.623  &
LSTM    &  0.099  &
LSTM    &  0.099  &
LSTM    &  0.462  \\

Conv1D  &  0.338  &
Conv1D  &  0.191  &
Conv1D  &  0.620  &
Conv1D  &  0.134  &
Conv1D  &  0.065  &
Conv1D  &  0.471  \\

\bottomrule
\end{tabular}
\end{table*}
\end{center}

There are a number of future directions that we would like to explore. 
An important parameter of online \iot device identification is how often do we
need to retrain the models on the edge device. Training just one model for all
devices/categories can be done in tens of seconds. Therefore, training for
this type of models is not prohibitive and can be done often, for example during
the night when the \iot gateway is less loaded due to lower network traffic. On
the other hand, frequent updating of models for each device and category might
not be feasible if the respective home \iot network has a considerable number of
devices and categories. In our case, having tens of devices means that
training a separate model for each device and category would take approximately
four hours in the best case scenario. Retraining a separate model for each category
would take approximately five minutes for six categories. As we have shown, the
time it takes to retrain the models depends on the type of neural networks, and
which part of the network is retrained, \ie  whether part of the model is
frozen or not. There is a trade-off between the decay of model accuracy per day
and the computation incurred by retraining the model. 


In a scenario where a user connects a new device to their
home network, we investigate and evaluate whether it is possible to infer the
device category. For that purpose we trained all our models of different types and training
lengths with data omitting exactly \emph{one} \iot device. This process was
repeated for every device in our test-bed. This led to training of more than 158,000 ML models. These models were evaluated using the data
collected from the omitted \iot device.

Table~\ref{tab:categorisation} shows the average \fscore over all window sizes
for each model type.  We see that different models perform
similarly and there is no major outlier. The lowest score was achieved for the
appliance and the hubs category. Surprisingly, the media category achieved a
rather low score as well. On the other hand, surveillance and home-automation
category achieved on average a rather high \fscore of $40\%$ and $44\%$
respectively. The highest score was achieved by the audio category ($62\%$).
However, this category contained only five devices, four of which were from the
same manufacturer (Amazon Alexa device). Therefore, these devices have very
similar network traffic which can lead to high category classification even if a
device is omitted from the training set.

\takeaway{Our results show that accurate inference of the category of a newly
connected device is hard and is an important direction for further research.}

While we have explored the influence of freezing different parts of the model and
its implications on the \fscore of the model, we evaluated only the average
change over all devices. It is possible that freezing of layers influence
different devices in different way. Therefore, freezing of different numbers of
layers depending on the device should be explored. 

While a single model for all devices/categories can be trained faster and
achieves higher \fscore, given that every household is essentially unique with a
different number and types of devices, it is impossible to create a model for
every permutation of \iot devices. On the other hand, having a separate model
for each device increases the training, as well as inference time. It also
achieves lower \fscore.  Therefore, we plan to investigate the possibility of
merging retrained binary classification models into a single
multi-classification model at the edge. 

Because we focus on online traffic classification, we rely on features that can
be readily extracted from the current network flow. Other approaches, where
historical data are included, \eg number of IPs contacted in the last hour or
DNS requests made, might be more reliable and lead to higher \fscore. Therefore,
we plan to investigate whether adding historical data as feature improves
the classification accuracy.

We focus on three most common neural networks which are widely popular. We
mostly use the same type of layer in the whole model in order to evaluate
influence of the specific type of the network on the training and inference time,
as well as on the \fscore. A combination of several layer types might lead to
higher classification accuracy.

We have shown that a single device might behave differently depending on the
other devices in the network. This was demonstrated when a model trained on one
test-bed achieved very low \fscore on the other test-bed and vice versa.
Therefore we would like to study dynamics of the networks and how the network
profile of devices changes depending on adding or removing other devices from
the network. 

One possible fallacy of model retraining is the possibility of malicious devices
being brought into the network. When retraining takes place, these might
effectively ``poison'' the model, allowing their malicious behavior to go
undetected.  However, this can be mitigated through adding signatures or
obfuscated code on the device~\cite{2019IoTGuardNDSS}, and it is out of scope
for this work.

\section{Conclusion}
\label{s:conclusions}

In this paper, we trained and evaluated over 200,000 different ML models for IoT device fingerprinting using full packet samples from a large number of IoT devices and categories, in active and idle mode.
We showed that the accuracy of the model decays over time, irrespective of the size of the training set. 
A trained model used the following day achieves on average $78\%$ accuracy. 
This accuracy drops to $72\%$ after a week and only $52\%$ after two weeks of usage. 
We also showed that a model trained on one dataset performs poorly when tested on a different dataset even from the same test-bed.
The accuracy of a model trained on \emph{idle} dataset drops to only $30\%$ when evaluated on \emph{active} data from the same test-bed and staggering $15\%$ when evaluated on a different test-bed contains a subset of \iot devices.  
We also show that even though retraining a model on an \emph{active} dataset from one test-bed increases the accuracy on the said test-bed from $30\%$ to $78\%$, it has very little impact on the other test-bed, increasing accuracy from $15\%$ to only $33\%$. 
The similar results were obtained when models were retrained on the second test-bed and evaluated on the first one. Our results clearly demonstrate that models need to be regularly retrained locally at the edge.

To address these issues, we evaluated model retraining at the edge using a representative edge device (Raspberry Pi 4). 
We showed that it is feasible to update a globally trained model with local data and achieve comparable accuracy to the globally trained model. 
We also showed that it is possible to speed up the training by partially freezing the parts of a model, and evaluated the impact of
freezing on the training time (can be cut by more than $66\%$, depending on the model) and the accuracy of the model (drop by less than $2\%$).

Our results clearly indicate that creating one general model is not a feasible solution for efficient \iot device identification at the edge due to the accuracy decay over time. 
The solution for this problem is to keep the model updated with local data and perform regular model retraining at the edge.  
In this way, the home gateway is capable of performing online \iot device classification, while retraining the models regularly during idle periods.
Our findings can be a step towards accurate and real-time IoT anomaly detection and threat mitigation.
